\documentclass[
preprint,
 amsmath,amssymb,
pra,
]{revtex4-1}
\usepackage{dcolumn}
\usepackage{bm}
\usepackage{amsfonts}
\usepackage{amsmath}
\usepackage{amssymb}
\usepackage{float}
\usepackage{graphicx}
\usepackage{subfigure}
\usepackage{epstopdf}
\usepackage{hyperref}

\begin{document}

\title{Enhancement of Steady-state Bosonic Squeezing and Entanglement in a Dissipative Optomechanical System}
\author{Chang-Geng Liao$^{1,2,3}$}
\author{Hong Xie$^{1,2,4}$}
\author{Xiao Shang$^{1,2}$}
\author{Zhi-Hua Chen$^{1,2}$}
\author{Xiu-Min Lin$^{1,2}$}
\thanks{xmlin@fjnu.edu.cn}

\affiliation{$^{1}$ Fujian Provincial Key Laboratory of Quantum Manipulation and New Energy Materials, College of Physics and Energy, Fujian Normal University, Fuzhou 350117, China}
\affiliation{$^{2}$ Fujian Provincial Collaborative Innovation Center for Optoelectronic Semiconductors and Efficient Devices, Xiamen 361005, China}
\affiliation{$^{3}$ Department of Electronic Engineering, Fujian Polytechnic of Information Technology, Fuzhou, 350003, China}
\affiliation{$^{4}$ College of JinShan, Fujian Agriculture and Forestry University, Fuzhou 350002, China}

\begin{abstract}
  We systematically study the influence of amplitude modulation on the steady-state bosonic squeezing and entanglement in a dissipative three-mode optomechanical system, where a vibrational mode of the membrane is coupled to the left and right cavity modes via the radiation pressure. Numerical simulation results show that the steady-state bosonic squeezing and entanglement can be significantly enhanced by periodically modulated external laser driving either or both ends of the cavity. Remarkably, the fact that as long as one periodically modulated external laser driving either end of the cavities is sufficient to enhance the squeezing and entanglement is convenient for actual experiment, whose cost is that required  modulation period number for achieving system stability is more. In addition, we numerically confirm the analytical prediction for optimal modulation frequency and discuss the corresponding physical mechanism.

\end{abstract}

\maketitle

\section{introduction}

Driven by a variety of different goals and promising prospects, cavity optomechanics, a field at the intersection of nanophysics and quantum optics, has developed over the past few years \cite{marquardt2009trend,meystre2013short,aspelmeyer2014cavity}. It has been known that nonclassical states of macroscopic mechanical resonators, especially the squeezed and entangled states, play a key role in test of the fundamental principles of quantum mechanics, quantum information processing, and ultrahigh-precision measurements. Many researches have been investigated on quantum squeezing and entanglement generation in cavity optomechanical interfaces. Normally, one can simply use radiation pressure forces or combine continuous quantum measurements and feedback to obtain stationary squeezing \cite{woolley2008nanomechanical,nunnenkamp2010cooling,liao2011parametric} and stationary entanglement\cite{Vitali2007Optomechanical,Tian2012Parametric} in a two-mode optomechanical system. In order to increase the richness of the research, the three-mode optomechanical setting was introduced and has been realized experimentally recently \cite{hill2012coherent,dong2012optomechanical,massel2012multimode}. Several theoretical schemes for generating quantum squeezing and entanglement in the three-mode optomechanical system have been proposed based on the basic idea that the auxiliary mode mediates an effective two-mode squeezing interaction between the two target modes \cite{mancini2002entangling,paternostro2007creating,Barzanjeh2011Entangling,Sh2012Reversible,Genes2009Chapter}. However, the schemes are generally restricted to the requirement of stability so that they yield at best a relatively small amount of squeezing and entanglement.

Resent studies show that large degrees of squeezing and entanglement can be achieved by mildly modulating the amplitude of the driving field \cite{Mari2009Gently,mari2012opto,schmidt2012optomechanical,chen2014enhancement,abdi2015entangling,li2015generation} or combining with dissipation mechanism \cite{Gu2013Squeezing,wang2016macroscopic}, where no feedback is needed. Moreover, the modulation-assisted driving can give rise to interesting and rich quantum dynamics \cite{rogers2012entanglement,doria2011optimal}. Farace and Giovannetti \cite{farace2012enhancing} further investigated this modulation regime and showed that simultaneous modulations of the mechanical frequency and input laser intensity can either enhance or weaken the desired quantum effects. Newly, the robust entanglement is generated by modulating the coupling strength between two mechanical oscillators \cite{PhysRevA.97.022336,PhysRevA.97.042314}. Besides, the modulation-induced mechanical parametric amplification effectively enhances the resonant optomechanical interaction and leads to single-photon strong-coupling\cite{PhysRevA.95.053861}. Remarkably, several works \cite{wang2013reservoir,kronwald2013arbitrarily,woolley2014two,wang2015bipartite,chen2015dissipation,NJP14,Chen2017Dissipativegeneration} reveal that optimizing relative ratio of optomechanical couplings, rather than simply increasing their magnitudes, is essential for achieving strong steady-state squeezing and entanglement via dissipation mechanism. These schemes exploit the Bogoliubov-mode-based method\cite{PhysRevLett.110.233602} instead of the S$\phi$rensen-M$\phi$lmer approach\cite{PhysRevA.88.062341}. Another promising means for generating strong entanglement or squeezing is the phonon-mediated four-wave mixing process\cite{Strong squeezing2014}. Although the physical explanations for these schemes are not quite the same, a common feature is to induce an effective engineered reservoir by driving the optomechanical systems with proper blue and red detuned lasers\cite{Strong squeezing2014,PhysRevLett.110.233602,PhysRevA.88.062341,PhysRevA.79.024301,Chen:17}.

In this work, combination of the modulation and the dissipation is considered. We expand the optomechanical model in~\cite{Mari2009Gently} to three-mode optomechanical system, which is similar to that in~\cite{liao2015enhancement} and \cite{huan2015dynamic} except being driven by periodic modulation field. A single-cavity optomechanical system usually requires an external laser to drive the mechanical resonator out of its zero steady state at equilibrium position. For the system considered here, an external laser being applied to either end of the cavity is sufficient to drive the vibrating membrane. Numerical simulation results show that the squeezing and entanglement can be enhanced with one-end or two-end periodically modulated external laser.  The time required for the two-end modulation when the system  achieves a stable state is shorter than that for the one-end modulation, but the one-end modulation reduces the difficulty of the experiment. What is more, with the help of the third mode acted as an engineered reservoir, dissipation mechanism is explored. Compared to the previous studies of three-mode modulated optomechanics \cite{abdi2015entangling,li2015generation,wang2016macroscopic}, more general modulations of quantum dynamics are discussed here.

In what follows, we give a detailed description of our model and obtain the linearized dynamical equations for the system in Sec.~II. In Sec.~III, analytical solutions for mean values in the cases of symmetric and asymmetric modulation are obtained in a perturbative way. Then We analyze in detail the characters of the mean values, where the numerical results agree well with the analytical results. In Sec.~IV, the mechanisms of squeezing and entanglement via combinations of the periodic amplitude modulation and the dissipation regime are discussed by assuming a simple but justifiable form of the effective coupling. Finally, conclusions are presented in Sec.~V.

\section{Theoretical model}
\begin{figure}[t]
\centering
  {\includegraphics[width=7cm]{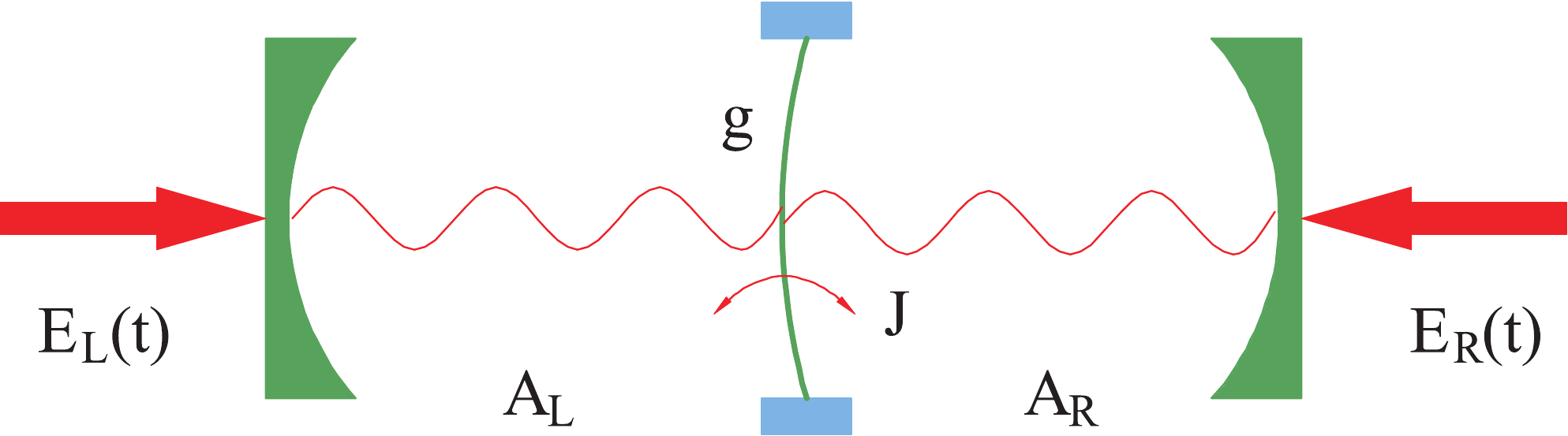}}
  \caption{\label{fig:mode}
 Schematic diagram of the optomechanical system.}
\end{figure}

The considered system is depicted in Fig.~\ref{fig:mode}. A dielectric membrane as a mechanical oscillator separates an optical cavity into two cavities and constructs a ``membrane-in-the-middle'' configuration, which has been theoretically studied  \cite{chen2014enhancement,bhattacharya2008optomechanical,jayich2008dispersive,miao2009standard,cheung2011nonadiabatic,biancofiore2011quantum,ludwig2012enhanced,stannigel2012optomechanical,teng2012quantum,PhysRevA.88.042331} and experimentally implemented  \cite{thompson2008strong,sankey2010strong,karuza2012optomechanical,purdy2012cavity,karuza2013optomechanically,karuza2012tunable,andrews2014bidirectional,lee2015multimode,purdy2015optomechanical}.
The mechanical oscillator with frequency ${\omega _m}$ is simultaneously coupled to the left and right cavity modes via the radiation pressure difference between the two cavities, where tunneling of photons through the membrane is allowed. The two cavity modes with frequency ${\omega _{{\rm{cL}}}}$ and ${\omega _{{\rm{cR}}}}$ are respectively driven by external lasers with periodically modulated amplitudes ${E_{\rm{L}}}\left( t \right)$ and ${E_{\rm{R}}}\left( t \right)$. In the rotating frame with respect to laser frequencies ${\omega _{\rm{L}}}$ and ${\omega _{\rm{R}}}$, the corresponding Hamiltonian reads ($\hbar  = 1$)
\begin{eqnarray}
H &= &\sum\limits_{j = {\rm{L,R}}} {[{\Delta _j}A_j^{\dag}{A_j}}  + i{E_j}(t)A_j^{\dag} - iE_j^ * (t){A_j}]  + \frac{{{\omega _{\rm{m}}}}}{2}({P^2} + {Q^2})\\\nonumber
&&+ g(A_{\rm{L}}^{\dag}{A_{\rm{L}}} - A_{\rm{R}}^{\dag}{A_{\rm{R}}})Q +J({A_{\rm{L}}}A_{\rm{R}}^ {\dag}+ A_{\rm{L}}^{\dag}{A_{\rm{R}}}).
\end{eqnarray}
Here, ${\Delta _j} = {\omega _{{\rm{c}}j}} - {\omega _j}$ denotes the $j$th cavity mode detuning, $A_j^{\dag}$ and ${A_j}$ represent the creation and annihilation operators of the $j$th cavity mode, $Q$ and $P$ are the dimensionless position and momentum operators of the mechanical mode with the standard canonical commutation relation $[Q,P] = i$, $J$ expresses the cavity-cavity coupling strength which is in the regime $J \ll {\omega _{{\rm{cL}}}},{\omega _{{\rm{cR}}}}$, and $g$ signifies the phonon-photon coupling coefficient. The time-dependent amplitude ${E_j}(t)$ is a period function with the period $\tau  $, i.e., ${E_j}(t + \tau ) = {E_j}(t)$.
Taken into account the cavity leakage and membrane damping, the dissipative dynamics of the system is described by the following nonlinear quantum Langevin equations (QLEs)
\begin{subequations}\label{L}
\begin{align}
\dot{Q} =&{\omega _m}P, \\
\dot{P} =&- {\omega _{\rm{m}}}Q - g(A_{\rm{L}}^{\dag}{A_{\rm{L}}} - A_{\rm{R}}^{\dag} {A_{\rm{R}}}) - {\gamma _{\rm{m}}}P + \xi (t), \\
\dot{A_{\rm{L}}}= &- (\kappa  + i{\Delta _{\rm{L}}}){A_{\rm{L}}} - ig{A_{\rm{L}}}Q - iJ{A_{\rm{R}}} + {E_{\rm{L}}}(t)+ \sqrt {2\kappa } a_{\rm{L}}^{{\rm{in}}}(t), \\
\dot{A_{\rm{R}}}= &- (\kappa  + i{\Delta _{\rm{R}}}){A_{\rm{R}}} + ig{A_{\rm{R}}}Q - iJ{A_{\rm{L}}} + {E_{\rm{R}}}(t)+ \sqrt {2\kappa } a_{\rm{R}}^{{\rm{in}}}(t),
\end{align}
\end{subequations}
where $\kappa $ and ${\gamma _{\rm{m}}}$ are severally the leakage rate of the cavities and the mechanical damping rate. The zero-mean fluctuation terms $a_j^{{\rm{in}}}(t)$ obey the correlation relations \cite{gardiner2004quantum}
\begin{subequations}\label{corr}
\begin{align}
&\left\langle {a_j^{{\rm{in}}}(t)a_j^{{\rm{in \dag }}}({t^\prime })} \right\rangle  = ({\overline n _{\rm{a}}} + 1)\delta (t - {t^\prime }), \\
&\left\langle {a_j^{{\rm{in \dag }}}(t)a_j^{{\rm{in}}}({t^\prime })} \right\rangle  = {\overline n _{\rm{a}}}\delta (t - {t^\prime }),
\end{align}
\end{subequations}
where ${\overline n _{\rm{a}}}={[\exp ({{{\hbar\omega _{{\rm{c}}j}}} \mathord{\left/{\vphantom {{{\omega _{{\rm{c}}j}}} {{k_{\rm{B}}}}}} \right.\kern-\nulldelimiterspace} {{k_{\rm{B}}}}}T) - 1]^{ - 1}}$ is the mean bath photon number at the environmental temperature $T$. The correlation function of zero-mean Brownian motion noise operator $\xi (t)$ in the case of the large mechanical quality factor $\mathbb{Q}=\omega _{\rm{m}}/\gamma_{\rm{m}}\gg1$ can be approximately described by the Markovian process and satisfies
\begin{equation}\label{mark}
\langle {\xi (t)\xi ({t^\prime }) + \xi ({t^\prime })\xi (t)} \rangle/2  = {\gamma _{\rm{m}}}(2{\overline n _{\rm{m}}} + 1)\delta (t - {t^\prime }),
\end{equation}
where $\overline n _{\rm{m}} = {[\exp ({{{\hbar\omega _{\rm{m}}}} \mathord{\left/{\vphantom {{{\omega _{\rm{m}}}} {{k_{\rm{B}}}}}} \right.\kern-\nulldelimiterspace} {{k_{\rm{B}}}}}T) - 1]^{ - 1}}$ is the mean thermal phonon number at the environmental temperature $T$.

In the presence of strong external driving fields, we can rewrite each Heisenberg operator as $O = \langle {O(t)}\rangle  + o{\rm{ }}$ $(O = Q,P,{A_j})$, where $o$ is quantum fluctuation operator around classical $c$-number mean value $\langle {O(t)}\rangle$. After applying standard linearization technique to the Eq.~(\ref{L}), we obtain the equations for the mean values
\begin{subequations}\label{mean}
\begin{align}
\dot{\langle Q \rangle} =&{\omega _{\rm{m}}}\langle P \rangle, \\
\dot{\langle P \rangle} =&- {\omega _{\rm{m}}}\langle Q \rangle  - {\gamma _{\rm{m}}}\langle P \rangle  - g({\langle {{A_{\rm{L}}}} \rangle ^ * }\langle {{A_{\rm{L}}}} \rangle - {\langle {{A_{\rm{R}}}} \rangle ^ * }\langle {{A_{\rm{R}}}} \rangle ), \\
\dot{\langle {{A_{\rm{L}}}} \rangle}=&- {\rm{(}}\kappa  + i{\Delta _{\rm{L}}}{\rm{)}}\langle {{A_{\rm{L}}}}\rangle  - ig\langle {{A_{\rm{L}}}} \rangle \langle Q\rangle - iJ\langle {{A_{\rm{R}}}} \rangle  + {E_{\rm{L}}}(t), \\
\dot{\langle{{A_{\rm{R}}}} \rangle}=&- {\rm{(}}\kappa  + i{\Delta _{\rm{R}}}{\rm{)}}\langle {{A_{\rm{R}}}} \rangle {\rm{ + }}ig\langle {{A_{\rm{R}}}} \rangle\langle Q \rangle- iJ\langle {{A_{\rm{L}}}} \rangle  + {E_{\rm{R}}}(t),
\end {align}
\end{subequations}
the linearized QLEs for the quantum fluctuations
\begin{subequations}\label{fmean}
\begin{align}
\dot{q}=&\omega {}_{\rm{m}}p, \\
\dot{p}=& - \omega {}_{\rm{m}}q - {\gamma _{\rm{m}}}p - g({\left\langle {{A_{\rm{L}}}} \right\rangle ^ * }{a_{\rm{L}}} - {\left\langle {{A_{\rm{R}}}} \right\rangle ^ * }{a_{\rm{R}}} + {\rm{h}}{\rm{.c}}{\rm{.}}) + \xi (t), \\
\dot a_{\rm{L}}= &- (\kappa  + i{\Delta _{\rm{L}}}){a_{\rm{L}}} - ig(\left\langle {{A_{\rm{L}}}} \right\rangle q{\rm{ + }}\left\langle Q \right\rangle {a_{\rm{L}}}) - iJ{a_{\rm{R}}}+ \sqrt {2\kappa } a_{\rm{L}}^{{\rm{in}}}(t), \\
\dot a_{\rm{R}}= &- (\kappa  + i{\Delta _{\rm{R}}}){a_{\rm{R}}} + ig(\left\langle {{A_{\rm{R}}}} \right\rangle q{\rm{ + }}\left\langle Q \right\rangle {a_{\rm{R}}})  - iJ{a_{\rm{L}}}+ \sqrt {2\kappa } a_{\rm{R}}^{{\rm{in}}}(t),
\end{align}
\end{subequations}
and the corresponding linearized system Hamiltonian
\begin{eqnarray}\label{lH}
{H^{{\rm{lin}}}}&=&({\Delta _{\rm{L}}} + g\langle Q \rangle )a_{\rm{L}}^ {\dag}{a_{\rm{L}}} + ({\Delta _{\rm{R}}} - g\langle Q \rangle )a_{\rm{R}}^ {\dag} {a_{\rm{R}}}+ \frac{{{\omega _{\rm{m}}}}}{2}\times({p^2} + {q^2}) + J(a_{\rm{L}}^ {\dag} {a_{\rm{R}}} + a_{\rm{R}}^ {\dag} {a_{\rm{L}}})\nonumber\\
&&+ g({\langle {{A_{\rm{L}}}} \rangle ^ * }{a_{\rm{L}}}+ \langle {{A_{\rm{L}}}} \rangle a_{\rm{L}}^{\dag} -{\langle {{A_{\rm{R}}}} \rangle ^ * }{a_{\rm{R}}} -  \langle {{A_{\rm{R}}}} \rangle a_{\rm{R}}^{\dag})q.
\end{eqnarray}

\section{The characters of the mean values}
It is difficult to find exact solutions of the mean values in Eq.~(\ref{mean}) in general. But when the system is far away from optomechanical instabilities and multistabilities \cite{ludwig2008optomechanical}, the optomechanical coupling can be treated in a perturbative way. More specifically, approximately analytical solutions of the mean values can be found by expanding them in power series of the coupling costant $g$. Besides, it is justifiable that stable solution has the same periodicity $\tau$ as the implemented modulation field ${E_j}(t)$. Hence, we can perform double expansions for the mean values $\langle {O(t)}\rangle$ in power series of $g$ and Fourier series, i.e.,
\begin{equation}\label{Ocoef}
\left\langle {O(t)} \right\rangle  = \sum\limits_{l = 0}^\infty  {\sum\limits_{n =  - \infty }^\infty  {{O_{n,l}}{e^{in\Omega t}}{g^l}} },
\end{equation}
where $\Omega  = {{2\pi } \mathord{\left/
 {\vphantom {{2\pi } \tau }} \right.
 \kern-\nulldelimiterspace} \tau }$ is the fundamental modulation frequency.
Similarly, Fourier series for the periodic driving amplitudes can be written as
\begin{subequations}\label{Ecoef}
\begin{align}
&\ {E_{\rm{L}}}(t) = \sum\limits_{n =  - \infty }^\infty  {E_n^{\rm{L}}{e^{  in\Omega t}}}, \\
&\ {E_{\rm{R}}}(t) = \sum\limits_{n =  - \infty }^\infty  {E_n^{\rm{R}}{e^{ in\Omega t}}}.
\end{align}
\end{subequations}
After directly substituting Eqs.~(\ref{Ocoef}) and~(\ref{Ecoef}) into Eq.~(\ref{mean}), the coefficients ${O_{n,l}}$ are completely determined by the following relations
\begin{subequations}\label{coef0}
\begin{align}
&\ {P_{n,0}} = {Q_{n,0}} = 0, \\
&\ A_{n,0}^{\rm{L}} = \frac{{iJE_{  n}^{\rm{R}} - (\kappa  + i{\Delta _{\rm{R}}} + in\Omega )E_{  n}^{\rm{L}}}}{{ - {J^2} - (\kappa  + i{\Delta _{\rm{R}}} + in\Omega )(\kappa  + i{\Delta _{\rm{L}}} + in\Omega )}},\\
&\ A_{n,0}^{\rm{R}} = \frac{{iJE_{  n}^{\rm{L}} - (\kappa  + i{\Delta _{\rm{L}}} + in\Omega )E_{ n}^{\rm{R}}}}{{ - {J^2} - (\kappa  + i{\Delta _{\rm{R}}} + in\Omega )(\kappa  + i{\Delta _{\rm{L}}} + in\Omega )}}
\end{align}
\end{subequations}
corresponding to the $0$-order perturbation with respect to $g$, and
\begin{subequations}\label{coef1}
\begin{align}
\ {P_{n,l}} =& \frac{{in\Omega }}{{{\omega _{\rm{m}}}}}{Q_{n,l}}, \\
\ {Q_{n,l}} = & - {\omega _{\rm{m}}}(\sum\limits_{k = 0}^{l - 1} {\sum\limits_{m =  - \infty }^\infty  {\frac{{A_{m,k}^{{\rm{L}} * }A_{n + m,l - k - 1}^{\rm{L}}}}{{\omega _{\rm{m}}^2 + i{\gamma _{\rm{m}}}n\Omega  - {{(n\Omega )}^2}}}} }- \sum\limits_{k = 0}^{l - 1} {\sum\limits_{m =  - \infty }^\infty  {\frac{{A_{m,k}^{{\rm{R}} * }A_{n + m,l - k - 1}^{\rm{R}}}}{{\omega _{\rm{m}}^2 + i{\gamma _{\rm{m}}}n\Omega  - {{(n\Omega )}^2}}}} } ),\\
\ A_{n,l}^{\rm{L}} = & - i\sum\limits_{k = 0}^{l - 1} {\sum\limits_{m =  - \infty }^\infty  {\frac{{A_{m,k}^{\rm{L}}{Q_{n - m,l - k - 1}} + JA_{n,l}^{\rm{R}}}}{{\kappa  + i{\Delta _{\rm{L}}} + in\Omega }}} },\\
\ A_{n,l}^{\rm{R}} = &i\sum\limits_{k = 0}^{l - 1} {\sum\limits_{m =  - \infty }^\infty  {\frac{{A_{m,k}^{\rm{R}}{Q_{n - m,l - k - 1}} - JA_{n,l}^{\rm{L}}}}{{\kappa  + i{\Delta _{\rm{R}}} + in\Omega }}} }
\end{align}
\end{subequations}
corresponding to the $l$-order coefficients in a recursive way.

In the case of identical cavity detuning ($\Delta  = {\Delta _{\rm{L}}} = {\Delta _{\rm{R}}}$) and symmetric modulation of the external driving laser [${E_{\rm{L}}}(t) = {E_{\rm{R}}}(t)$], it is reasonable to expect that the mean values $\left\langle {{A_{\rm{L}}}} \right\rangle$ and $\left\langle {{A_{\rm{R}}}} \right\rangle$ have the same stable solutions. Thus, Eqs.~(\ref{coef0}) and (\ref{coef1}) can be further simplified as follows:
\begin{subequations}\label{ana}
\begin{align}
{P_{n,0}} &= {Q_{n,0}} = 0,\\
A_{n,0}^{\rm{L}} &=A_{n,0}^{\rm{R}} =\frac{E_{n}^{\rm{L}}}{[\kappa+i(\Delta  + n\Omega +J)]}=\frac{E_{n}^{\rm{R}}}{[\kappa+i(\Delta  + n\Omega +J)]},\\
{P_{n,l}} &= {Q_{n,l}} =A_{n,l}^{\rm{L}}=A_{n,l}^{\rm{R}}= 0.
\end{align}
\end{subequations}

\begin{figure}[t]
 \centering
  \subfigure{\includegraphics[width=0.4\textwidth]{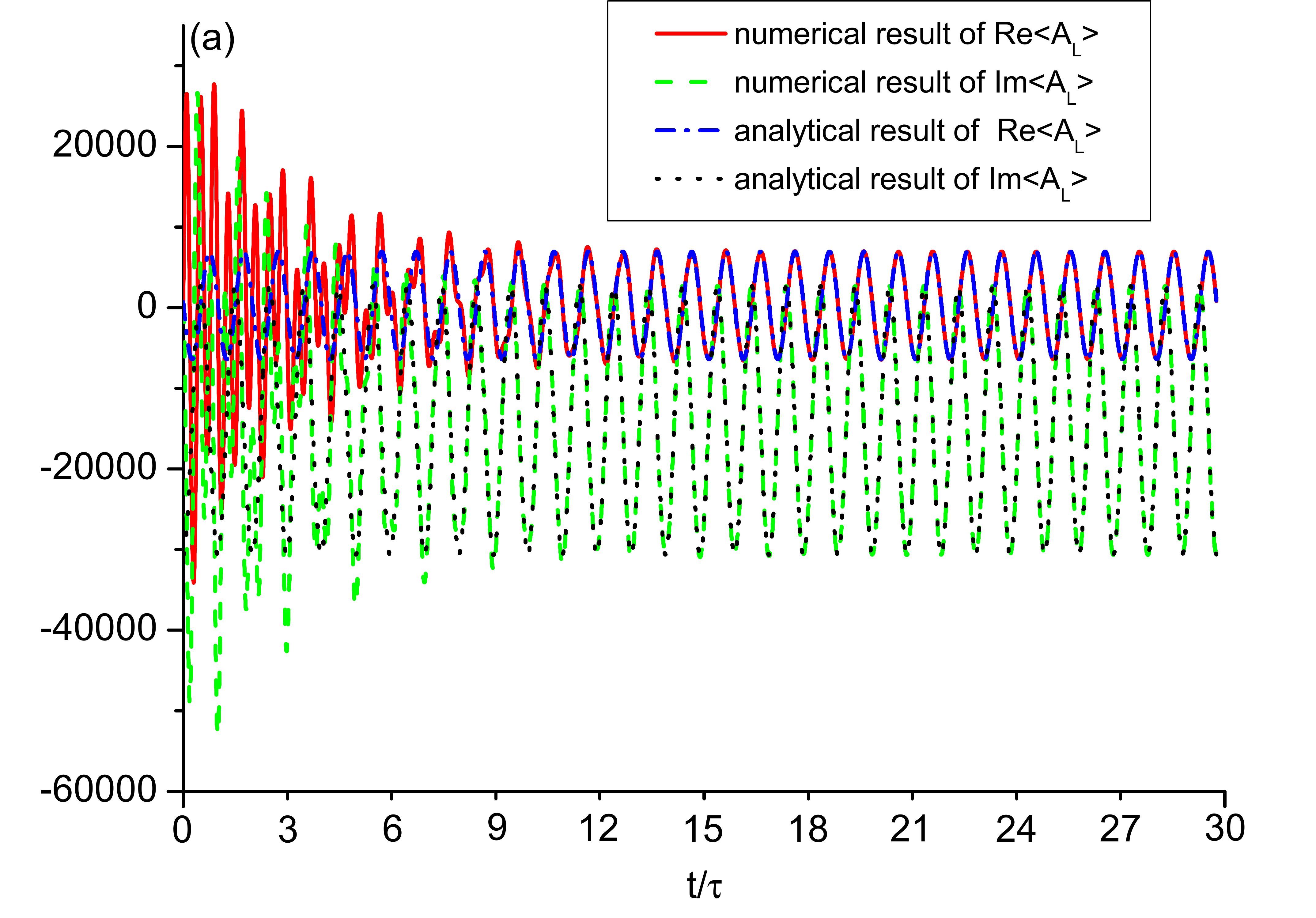}\label{fig:meana}}
    \subfigure{\includegraphics[width=0.4\textwidth]{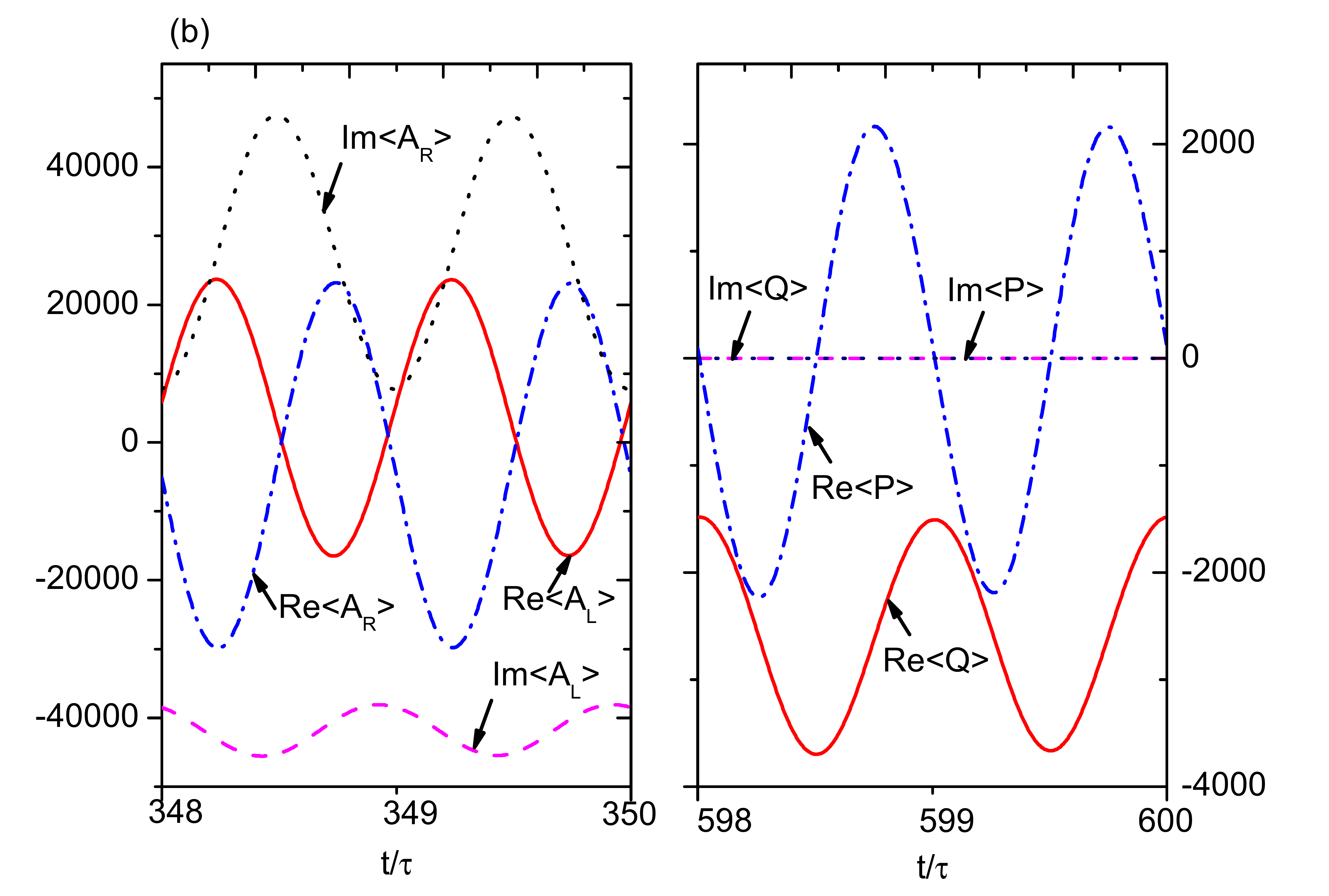}\label{fig:meanb}}
   \subfigure{\includegraphics[width=0.4\textwidth]{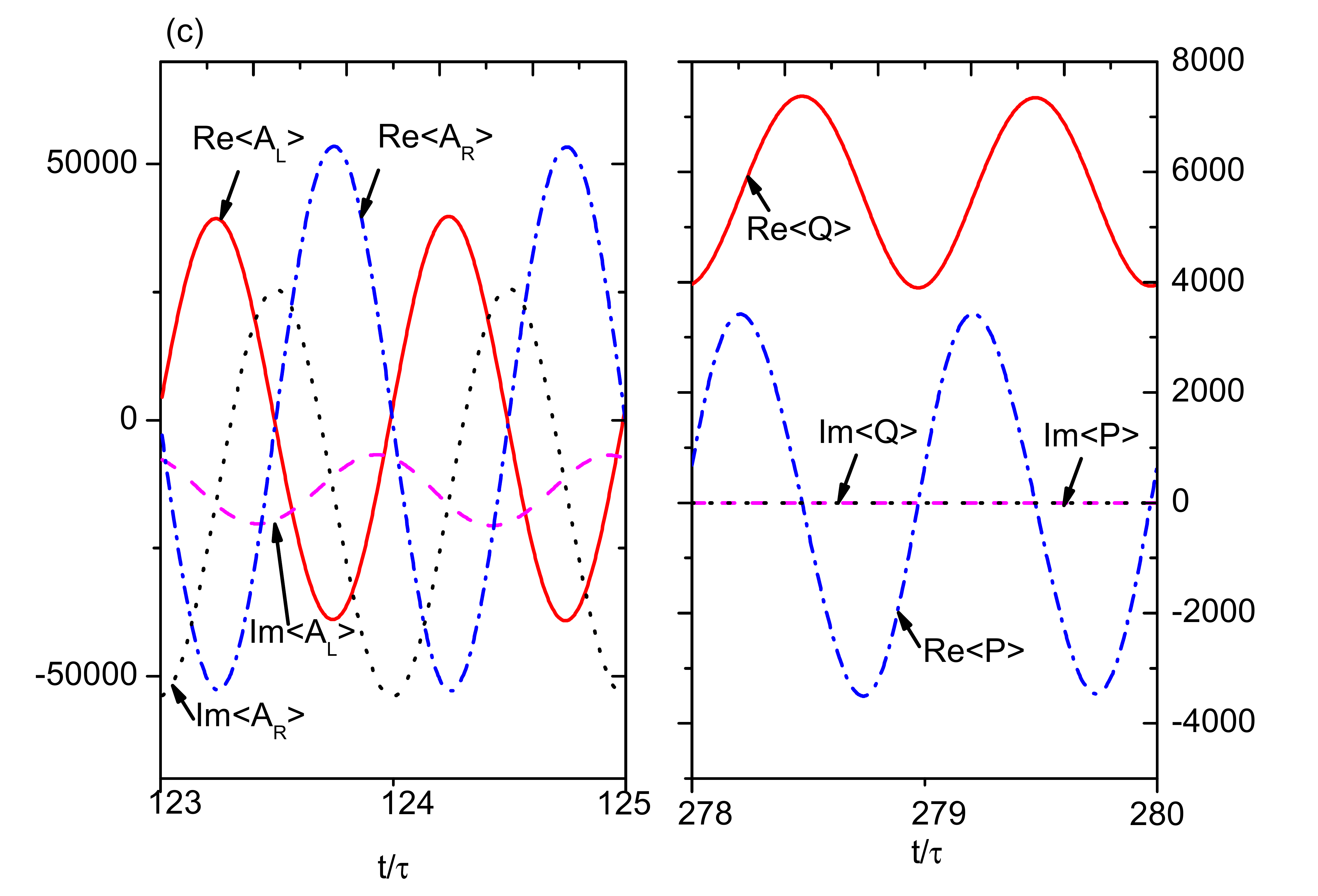}\label{fig:meanc}}
 \centering  \caption{\label{fig:mean} Time evolution of the real and imaginary parts of the mean values in the case of identical cavity detuning for three different modulation driving lasers. (a) symmetric modulation with ${E_{\rm{L}}}(t) = {E_{\rm{R}}}(t) = 7 \times {10^4} + 3.5 \times {10^4}\times{e^{ - i\Omega t}} + 3.5 \times {10^4}\times{e^{i\Omega t}}$; (b) single cavity driving with ${E_{\rm{L}}}(t) = 7 \times {10^4} + 3.5 \times {10^4}\times{e^{ - i\Omega t}} + 3.5 \times {10^4}\times{e^{i\Omega t}}$, ${E_{\rm{R}}}(t) = 0$; (c) single cavity modulation with ${E_{\rm{L}}}(t) = 7 \times {10^4} + 7 \times {10^4}\times{e^{ - i\Omega t}} + 7 \times {10^4}\times{e^{i\Omega t}}$, ${E_{\rm{R}}}(t) = 7 \times {10^4}$. The chosen parameters in units of ${\omega _{\rm{m}}}$ are: $\Omega  = 2$, $\kappa  = 0.1$, ${\gamma _{\rm{m}}} = 0.001$, $J = 2$, $\Delta  = 3$, and $g = 4 \times {10^{ - 6}}$.}
\end{figure}

Figure~\ref{fig:mean} gives both the numerical and the analytical results of the real and imaginary parts of the mean values in the case of identical cavity detuning for three different modulation driving laser. The numerical solutions of mean values corresponding to Eq.~(\ref{mean}) agree well with the analytical results of Eqs.~(\ref{Ocoef}) and (\ref{ana}) in the long time limit. Figure~\ref{fig:meana} displays the asymptotic evolution of the real
and imaginary parts of the left (or right) cavity mode mean value $\left\langle {{A_{\rm{L}}}} \right\rangle$ (or $\left\langle {{A_{\rm{R}}}} \right\rangle$) in the case of symmetric modulation driving laser. It is obvious that the numerical results (solid red and dashed green lines) agree well with the analytical results (dash dotted blue and dotted black lines) after about $15$ modulation periods. And the numerical results of the mean values $\langle P\rangle$ and $\langle Q \rangle$ obtained by Eq. (\ref{mean}) equal to zero, which are completely consistent with the analytical results of Eqs.~(\ref{Ocoef}) and (\ref{ana}) [no shown in Fig.~\ref{fig:meana}]. In Figs.~\ref{fig:meanb} and~\ref{fig:meanc}, we plot the the asymptotic evolution of the mean values in cases of single cavity driving and single cavity modulation, respectively. Since our calculations reveal that the numerical results agree well with the analytical results after about hundreds of modulation periods, we only plot the numerical solutions in the long time limit in order to avoid confusion. All results in  Fig.~\ref{fig:mean} show that the asymptotic evolution periods of the mean values are indeed $\tau$, where we have truncated the series in Eq. (\ref{ana}) to the terms with subscript $\left| n \right| \le 1$. We also find that the real parts of the mean values $\left\langle P\right\rangle$ and $\left\langle Q\right\rangle$ are no longer zero in the cases of single cavity driving and single cavity modulation, and the needed number of modulation period to achieve stable result varies with modulation mechanisms and parameters. For example, corresponding to three different modulation mechanisms and chosen parameters in Figs.~\ref{fig:meana},~\ref{fig:meanb}, and~\ref{fig:meanc}, the required numbers of modulation period for cavity modes to achieve the stable mean values respectively are about $15$, $348$, and $123$, while those for the mechanical oscillator are $598$ and $278$ [see Figs.~\ref{fig:meanb} and~\ref{fig:meanc}]. Obviously, from the point of the required time for obtaining steady state, the effect of symmetric modulation is the best.
\begin{figure}[t]
  \centering
  \subfigure{\includegraphics[width=0.4\textwidth]{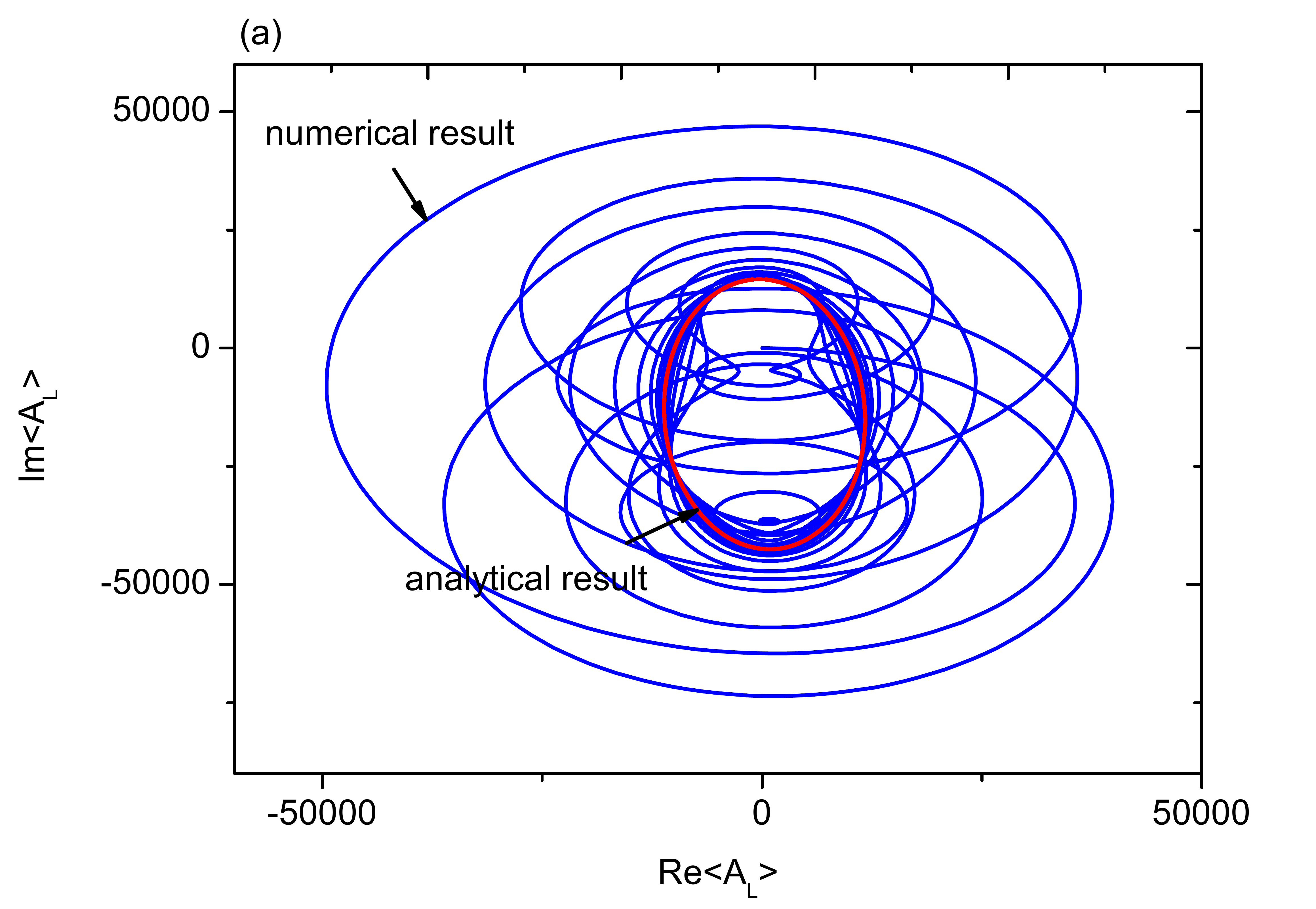}\label{fig:phasea}}
  \subfigure{\includegraphics[width=0.4\textwidth]{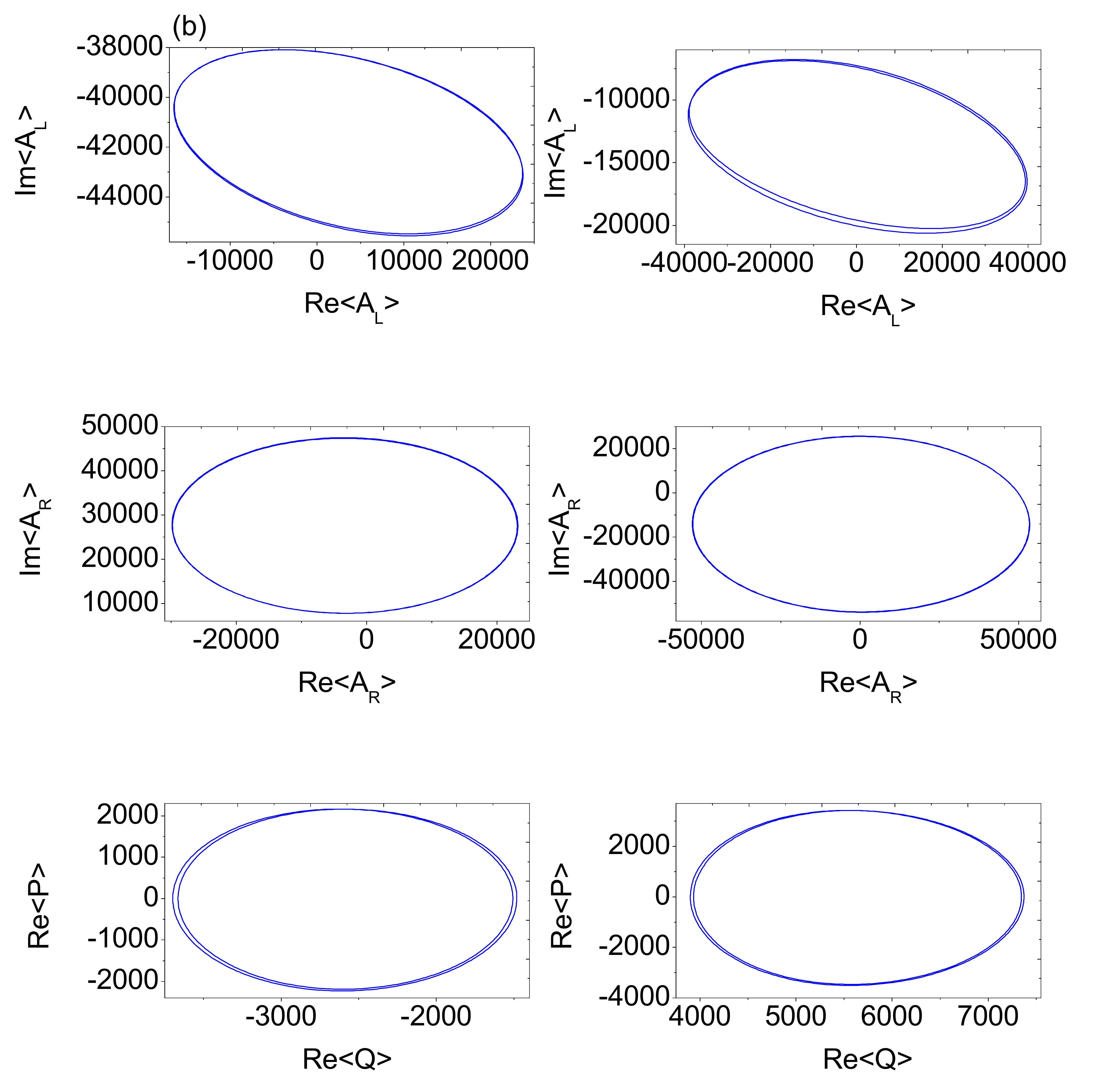}\label{fig:phaseb}}
  \caption{\label{fig:phase}
  Phase space trajectories of the classical $c$-number mean values. (a) Phase space trajectories of $\left\langle {{A_{\rm{L}}}(t)} \right\rangle$ from $t=0$ to $t=30\tau$ for symmetric modulation; (b) Phase space trajectories of cavity field mean values and the dimensionless mechanical position and momentum mean values for asymmetric modulation. The left and right columns are results of single cavity driving and single cavity modulation, respectively. All the chosen parameters are identical to those in Fig.~\ref{fig:mean}.}
\end{figure}
To gain more insights about the dynamics, we respectively plot the phase space trajectories of the mean values for symmetric and asymmetric modulations in Fig.~\ref{fig:phase}. As shown in Fig.~\ref{fig:phasea}, when the system is stable after dozens of modulation periods, the numerical phase space trajectories of $\left\langle {{A_{\rm{L}}}(t)} \right\rangle$ (or $\left\langle {{A_{\rm{R}}}(t)} \right\rangle$) finally converge to a limit cycle in the case of symmetric modulation, which agrees well with analytical prediction. In the cases of single cavity driving and single cavity modulation, the numerical results in Fig.~\ref{fig:phaseb} display that the phase space trajectories of the cavity mode mean values $\left\langle {{A_{\rm{L}}}(t)} \right\rangle$,$\left\langle {{A_{\rm{R}}}(t)} \right\rangle$, and the dimensionless mechanical position and momentum mean values almost converge to a limit cycle after hundreds of modulation periods.

\section{Stationary bosonic squeezing and entanglement}

Since the asymptotic evolution period of the system is $\tau $, without loss of generality, we assume the asymptotic form for time-dependent mean values of the cavity modes as follows:
\begin{subequations}\label{Aas}
\begin{align}
&\left\langle {{A_{\rm{L}}}(t)} \right\rangle  = {A_{{\rm{L}}0}} + {A_{{\rm{L1}}}}{e^{ - i\Omega t}}, \\
&\left\langle {{A_{\rm{R}}}(t)} \right\rangle  = {A_{{\rm{R0}}}} + {A_{{\rm{R1}}}}{e^{ - i\Omega t}},
\end{align}
\end{subequations}
where ${A_{j0}}$ and ${A_{j1}}$ are positive real number and related to the driving amplitude components $E_n^{\rm{L}}$ and $E_n^{\rm{R}}$ in Eq.~(\ref{Ecoef}).
When $t \to \infty$ and ${\omega _{\rm{m}}} \gg {\gamma _{\rm{m}}} > 0$, the corresponding mechanical mean values and the driving amplitude can be readily derived from Eq.~(\ref{mean}) via Laplace transformation and inverse transformation
\begin{subequations}\label{lap}
\begin{align}
\left\langle {P(t)} \right\rangle  \simeq &\frac{{ig\Omega ({A_{{\rm{R0}}}}{A_{{\rm{R1}}}} - {A_{{\rm{L0}}}}{A_{{\rm{L1}}}})}}{{({\Omega ^2} - \omega _{\rm{m}}^{\rm{2}})}}({e^{ - i\Omega t}} - {e^{i\Omega t}}), \\
\left\langle {Q(t)} \right
\rangle  \simeq &\frac{{g(A_{{\rm{R0}}}^2 + A_{{\rm{R1}}}^2 - A_{{\rm{L0}}}^{\rm{2}} - A_{{\rm{L1}}}^{\rm{2}})}}{{{\omega _{\rm{m}}}}}  + \frac{{g({A_{{\rm{R0}}}}{A_{{\rm{R1}}}} - {A_{{\rm{L0}}}}{A_{{\rm{L1}}}})}}{{{\omega _{\rm{m}}}}}\times(1 - \frac{{{\Omega ^2}}}{{{\Omega ^2} - \omega _{\rm{m}}^2}})({e^{ - i\Omega t}} + {e^{i\Omega t}}),\\
 {E_{\rm{L}}}(t) \simeq &E_{\rm{0}}^{\rm{L}}{\rm{ + }}E_{\rm{1}}^{\rm{L}}{e^{ - i\Omega t}}{\rm{ + }}E_{ - 1}^{\rm{L}}{e^{i\Omega t}}{\rm{ + }}E_{\rm{2}}^{\rm{L}}{e^{ - 2i\Omega t}},\\
{E_{\rm{R}}}(t) \simeq &E_{\rm{0}}^{\rm{R}}{\rm{ + }}E_{\rm{1}}^{\rm{R}}{e^{ - i\Omega t}}{\rm{ + }}E_{ - 1}^{\rm{R}}{e^{i\Omega t}}{\rm{ + }}E_{\rm{2}}^{\rm{R}}{e^{ - 2i\Omega t}},
\end{align}
\end{subequations}
with the driving amplitude components
\begin{subequations}\label{lap1}
\begin{align}
{E_{\rm{0}}^{\rm{L}}{\rm}} = & (\kappa + i{\Delta _{\rm{L}}}){A_{{\rm{L0}}}} + iJ{A_{{\rm{R0}}}} + \frac{{i{g^2}{A_{{\rm{L0}}}}(A_{{\rm{R0}}}^2 + A_{{\rm{R1}}}^{\rm{2}} - A_{{\rm{L0}}}^2 - A_{{\rm{L1}}}^{\rm{2}})}}{{{\omega _m}}} \\\nonumber
&+ \frac{{i{g^2}{A_{{\rm{L1}}}}({A_{{\rm{R0}}}}{A_{{\rm{R1}}}} - {A_{{\rm{L0}}}}{A_{{\rm{L1}}}})}}{{{\omega _{\rm{m}}}}}(1 - \frac{{{\Omega ^2}}}{{{\Omega ^2} - \omega _{\rm{m}}^2}}), \\
{E_{\rm{1}}^{\rm{L}}}=&[\kappa  + i({\Delta _{\rm{L}}} - \Omega )]{A_{{\rm{L1}}}} + iJ{A_{{\rm{R1}}}}+ \frac{{i{g^2}{A_{{\rm{L1}}}}(A_{{\rm{R0}}}^2 + A_{{\rm{R1}}}^2 - A_{{\rm{L0}}}^2 - A_{{\rm{L1}}}^{\rm{2}})}}{{{\omega _m}}}\\\nonumber
&+ \frac{{i{g^2}{A_{{\rm{L0}}}}({A_{{\rm{R0}}}}{A_{{\rm{R1}}}} - {A_{{\rm{L0}}}}{A_{{\rm{L1}}}})}}{{{\omega _{\rm{m}}}}}(1 - \frac{{{\Omega ^2}}}{{{\Omega ^2} - \omega _{\rm{m}}^2}}), \\
{E_{\rm{- 1}}^{\rm{L}}{\rm}} = &\frac{{i{g^2}{A_{{\rm{L0}}}}({A_{{\rm{R0}}}}{A_{{\rm{R1}}}} - {A_{{\rm{L0}}}}{A_{{\rm{L1}}}})}}{{{\omega _{\rm{m}}}}}(1 - \frac{{{\Omega ^2}}}{{{\Omega ^2} - \omega _{\rm{m}}^2}}),\\
{E_{\rm{2}}^{\rm{L}}{\rm}} = &\frac{{i{g^2}{A_{{\rm{L1}}}}({A_{{\rm{R0}}}}{A_{{\rm{R1}}}} - {A_{{\rm{L0}}}}{A_{{\rm{L1}}}})}}{{{\omega _{\rm{m}}}}}(1 - \frac{{{\Omega ^2}}}{{{\Omega ^2} - \omega _{\rm{m}}^2}}),\\
{E_{\rm{0}}^{\rm{R}}{\rm}} = &(\kappa  + i{\Delta _{\rm{R}}}){A_{{\rm{R0}}}} + iJ{A_{{\rm{L}}0}}- \frac{{i{g^2}{A_{{\rm{R0}}}}(A_{{\rm{R}}0}^2 + A_{{\rm{R1}}}^2 - A_{{\rm{L0}}}^2 - A_{{\rm{L1}}}^{\rm{2}})}}{{{\omega _{\rm{m}}}}}\\\nonumber
&- \frac{{i{g^2}{A_{{\rm{R1}}}}({A_{{\rm{R0}}}}{A_{{\rm{R}}1}} - {A_{{\rm{L0}}}}{A_{{\rm{L1}}}})}}{{{\omega _{\rm{m}}}}}(1 - \frac{{{\Omega ^2}}}{{{\Omega ^2} - \omega _{\rm{m}}^{\rm{2}}}}),\\
{E_{\rm{1}}^{\rm{R}}{\rm}}= &{\rm{[}}\kappa  + i({\Delta _{\rm{R}}} - \Omega )]{A_{{\rm{R1}}}} + iJ{A_{{\rm{L1}}}} - \frac{{i{g^2}{A_{{\rm{R1}}}}(A_{{\rm{R0}}}^2 + A_{{\rm{R1}}}^2 - A_{{\rm{L0}}}^2 - A_{{\rm{L1}}}^{\rm{2}})}}{{{\omega _m}}} \\\nonumber
&- \frac{{i{g^2}{A_{{\rm{R0}}}}({A_{{\rm{R0}}}}{A_{{\rm{R1}}}} - {A_{{\rm{L0}}}}{A_{{\rm{L1}}}})}}{{{\omega _{\rm{m}}}}}(1 - \frac{{{\Omega ^2}}}{{{\Omega ^2} - \omega _{\rm{m}}^2}}),\\
{E_{\rm{- 1}}^{\rm{R}} }= & - \frac{{i{g^2}{A_{{\rm{R}}0}}({A_{{\rm{R}}0}}{A_{{\rm{R}}1}} - {A_{{\rm{L}}0}}{A_{{\rm{L}}1}})}}{{{\omega _{\rm{m}}}}}(1 - \frac{{{\Omega ^2}}}{{{\Omega ^2} - \omega _{\rm{m}}^{\rm{2}}}}),\\
{E_{\rm{2}}^{\rm{R}}} =&  - \frac{{i{g^2}{A_{{\rm{R}}1}}({A_{{\rm{R}}0}}{A_{{\rm{R}}1}} - {A_{{\rm{L}}0}}{A_{{\rm{L}}1}})}}{{{\omega _{\rm{m}}}}}(1 - \frac{{{\Omega ^2}}}{{{\Omega ^2} - \omega _{\rm{m}}^2}}).
\end{align}
\end{subequations}

\begin{figure}[t]
\centering
  {\includegraphics[width=0.4\textwidth]{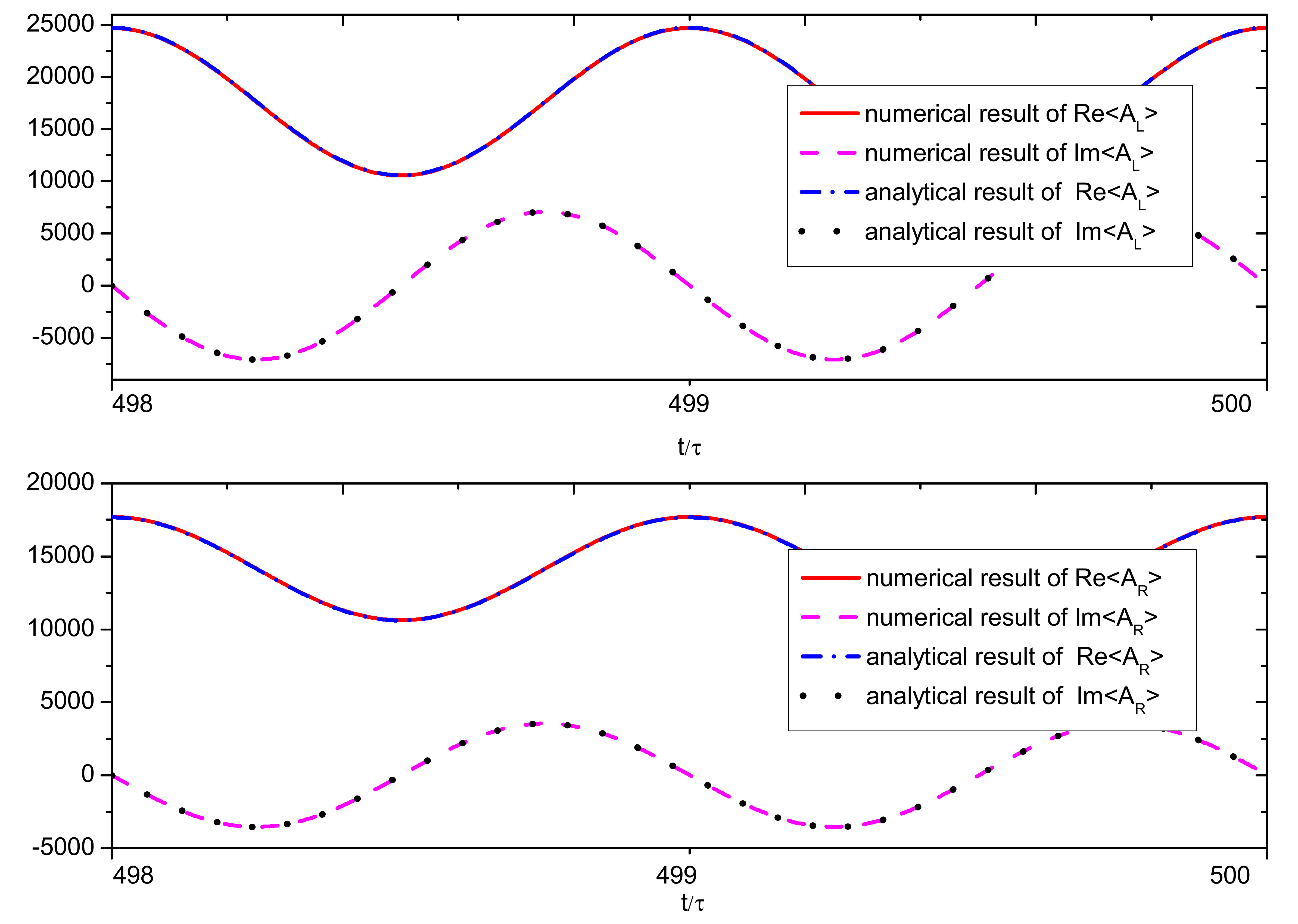}}
  \caption{\label{fig:cavitymean}
 Real and imaginary parts of cavity mode mean value $\left\langle {{A_j}(t)} \right\rangle$ as a function of time in the long time limit. The chosen parameters in units of ${\omega _{\rm{m}}}$ are: $\Omega  = 2$, $\kappa  = 0.1$, ${\gamma _{\rm{m}}} = 0.001$, $J = 2$, $\Delta  = 3$, $g = 4 \times {10^{ - 6}}$, ${A_{{\rm{L}}0}} = 0.1/{\sqrt{2}g}$, ${A_{{\rm{L}}1}} = 0.04/{\sqrt{2}g}$, ${A_{{\rm{R}}0}} = 0.08/{\sqrt{2}g}$, and ${A_{{\rm{R}}1}} = 0.02/{\sqrt{2}g}$.}
\end{figure}

In the long time limit, when driving amplitudes ${E_{\rm{L}}}(t)$ and ${E_{\rm{R}}}(t)$ with forms as Eqs.~(\ref{lap}) and (\ref{lap1}) are applied to Eq.~(\ref{mean}), Fig.~\ref{fig:cavitymean} numerically confirms that the time-dependent mean values of the cavity modes just as Eq.~(\ref{Aas}) are precisely generated, where the parameters ${A_{j0}}$ and ${A_{j1}}$ are taken as ${A_{{\rm{L}}0}} = 0.1/{\sqrt{2}g}$, ${A_{{\rm{L}}1}} = 0.04/{\sqrt{2}g}$, ${A_{{\rm{R}}0}} = 0.08/{\sqrt{2}g}$, and ${A_{{\rm{R}}1}} = 0.02/{\sqrt{2}g}$. In fact, the above four parameters can be arbitrary assigned when the requirement ${A_{{\rm{L1}}}} + {A_{{\rm{R1}}}} < {A_{{\rm{L0}}}} + {A_{{\rm{R0}}}}$ is met, which ensures stability. Thus, one can always design the corresponding modulation driving laser to realize mean values of the cavity modes with any
periodic form (the specific form is dependent on what effect we want to achieve).

In the following, based on the assumption of Eq.~(\ref{Aas}) we analyze how to enhance squeezing and entanglement via the symmetrically and asymmetrically periodic modulation.
By introducing the position and momentum quadratures for the two cavity modes and their input noises
\begin{subequations}
\begin{align}
&{x_j} = \frac{{{a_j} + a_j^ {\dag} }}{{\sqrt 2 }},\\
&{y_j} = \frac{{{a_j} - a_j^ {\dag} }}{{i\sqrt 2 }}, \\
&x_j^{{\rm{in}}}(t) = \frac{{a_j^{{\rm{in}}}(t) + a_j^{{\rm{in \dag }}}(t)}}{{\sqrt 2 }}, \\
&y_j^{{\rm{in}}}(t) = \frac{{a_j^{{\rm{in}}}(t) - a_j^{{\rm{in \dag }}}(t)}}{{i\sqrt 2 }},
\end{align}
\end{subequations}
and the column vectors of all quadratures and noises
\begin{subequations}
\begin{align}
U = &{(q,p,{x_{\rm{L}}},{y_{\rm{L}}},{x_{\rm{R}}},{y_{\rm{R}}})^{\rm{T}}},\\
N(t) =& (0,\xi (t),\sqrt {2\kappa } x_{\rm{L}}^{{\rm{in}}}(t),\sqrt {2\kappa } y_{\rm{L}}^{{\rm{in}}}(t),\sqrt {2\kappa } x_{\rm{R}}^{{\rm{in}}}(t),\sqrt {2\kappa } y_{\rm{R}}^{{\rm{in}}}(t))^{\rm{T}},
\end{align}
\end{subequations}
Eq.~(\ref{fmean}) can be rewritten as
\begin{eqnarray}\label{dU}
\dot{U} = {R(t)U + N(t)}
\end{eqnarray}
with
\begin{eqnarray}\label{R}
R(t) = \left( {\begin{array}{*{20}{c}}
0&{{\omega _{\rm{m}}}}&0&0&0&0\\
{ - {\omega _{\rm{m}}}}&{ - {\gamma _{\rm{m}}}}&{ - {G_{{\rm{Lr}}}}(t)}&{-{G_{{\rm{Li}}}}(t)}&{{G_{{\rm{Rr}}}}(t)}&{{G_{{\rm{Ri}}}}(t)}\\
{{G_{{\rm{Li}}}}(t)}&0&{ - \kappa }&{{\Delta _1}(t)}&0&J\\
{ - {G_{{\rm{Lr}}}}(t)}&0&{ - {\Delta _1}(t)}&{ - \kappa }&{ - J}&0\\
{ - {G_{{\rm{Ri}}}}(t)}&0&0&J&{ - \kappa }&{{\Delta _2}(t)}\\
{{G_{{\rm{Rr}}}}(t)}&0&{ - J}&0&{ - {\Delta _2}(t)}&{ - \kappa }
\end{array}}\right),
\end{eqnarray}
where the effective time-modulated detuning
 \begin{subequations}\label{effdelta}
\begin{align}
{\Delta _1}(t) = {\Delta _{\rm{L}}} + g\left\langle Q \right\rangle,\\
{\Delta _2}(t) = {\Delta _{\rm{R}}} - g\left\langle Q \right\rangle,
\end{align}
\end{subequations}
${G_{j{\rm{r}}}}(t)$ and ${G_{j{\rm{i}}}}(t)$ are respectively real and imaginary parts of the effective coupling coefficient
\begin{eqnarray}\label{effG}
{G_j}(t)=\sqrt 2 g\left\langle {{A_j}(t)} \right\rangle  =\sqrt 2 g({A_{j0}} + {A_{j1}}{e^{ - i\Omega t}}) ={G_{j0}} + {G_{j1}}{e^{ - i\Omega t}}.
\end{eqnarray}
When the system is stable, it converges to a time-dependent Gaussian state \cite{weedbrook2012gaussian}, which is independently from the initial condition. Thus, the asymptotic state of the fluctuation is fully described by the covariance matrix (CM) $\sigma (t)$ of the pairwise correlation among the quadratures, where the entries of the CM are defined as
\begin{eqnarray}\label{Ue}
\sigma_{k,l}=<U_k(t)U_l(t)+U_l(t)U_k(t)>/2.
\end{eqnarray}
From Eqs.~(\ref{dU}) and~(\ref{Ue}), it can be deduced
\begin{equation}\label{sig}
\dot{\sigma} (t)= R(t)\sigma (t) + \sigma (t)R{(t)^{\rm{T}}} + D,
\end{equation}
where $D$ is a diffusion matrix whose components are associated with the noise correlation functions and defined as
\begin{equation}
\delta (t - {t'}){D_{k,l}} = {{\left\langle {{N_k}(t)N_l^{\dag} ({t'}) + N_l^{\dag} ({t'}){N_k}(t)} \right\rangle } \mathord{\left/ {\vphantom {{\left\langle {{N_k}(t)N_l^{\dag} ({t'}) + N_l^{\dag} ({t'}){N_k}(t)} \right\rangle } 2}} \right.
 \kern-\nulldelimiterspace} 2}.
\end{equation}
It can be gained from Eqs.~(\ref{corr}) and (\ref{mark})
\begin{eqnarray}
D = {\rm{diag}}(0,{\gamma _{\rm{m}}}(2{\overline n _{\rm{m}}} + 1),\kappa (2{\overline n _{\rm{a}}} + 1),\kappa (2{\overline n _{\rm{a}}} + 1),\kappa (2{\overline n _{\rm{a}}} + 1),\kappa (2{\overline n _{\rm{a}}} + 1)).
\end{eqnarray}

In the long time limit, based on Floquet's theorem \cite{Mari2009Gently,mari2012opto,chen2014enhancement,teschl2012ordinary} the periodicity of the entries of $R(t)$ implies that asymptotic solution of the linear differential Eq.~(\ref{sig}) will have the same period $\tau$, i.e.,
\begin{equation}
\sigma (t) = \sigma (t + \tau ).
\end{equation}
The CM $\sigma (t)$ can be written as a block matrix
\begin{equation}
\sigma (t) = \left( {\begin{array}{*{20}{c}}
{{\sigma _{\rm{M}}}}&{{\sigma _{{\rm{ML}}}}}&{{\sigma _{{\rm{MR}}}}}\\
{\sigma _{{\rm{ML}}}^{\rm{T}}}&{{\sigma _{\rm{L}}}}&{{\sigma _{{\rm{LR}}}}}\\
{\sigma _{{\rm{MR}}}^{\rm{T}}}&{\sigma _{{\rm{LR}}}^{\rm{T}}}&{{\sigma _{\rm{R}}}}
\end{array}} \right),
\end{equation}
where each block represents a $2 \times 2$ matrix. The diagonal blocks represent the variance within each subsystem (for example,  resonator ${\rm{M}}$, the left cavity mode ${\rm{L}}$, and the right cavity mode ${\rm{R}}$), while the off-diagonal blocks denote covariance across different subsystems.
Since the asymptotic state of the system is Gaussian, it is convenient to measure the pairwise entanglement ${E_{\rm{N}}}$ with the logarithmic negativity\cite{vidal2002computable,adesso2004extremal}, which can be readily computed from the reduced $4 \times 4$ CM ${\sigma _{\rm{r}}}(t)$ for two subsystems
\begin{equation}
{\sigma _{\rm{r}}}(t) = \left( {\begin{array}{*{20}{c}}
{{\sigma _1}}&{{\sigma _{\rm{c}}}}\\
{\sigma _{\rm{c}}^{\rm{T}}}&{{\sigma _2}}
\end{array}} \right).
\end{equation}
The logarithmic negativity ${E_{\rm{N}}}$ is then given by
\begin{equation}
{E_N} = \max [0, - \ln (2\eta )]
\end{equation}
with
\begin{equation}
\eta \equiv 2^{-1/2}\{\Sigma-[\Sigma^2-4\det {\sigma _{\rm{r}}}]^{1/2}\}^{1/2},
\end{equation}
where
\begin{equation}
\Sigma \equiv \det {\sigma _1} + \det {\sigma _1} - 2\det {\sigma _{\rm{c}}}.
\end{equation}

\begin{figure}[t]
   \centering
   \subfigure{\includegraphics[width=0.4\textwidth]{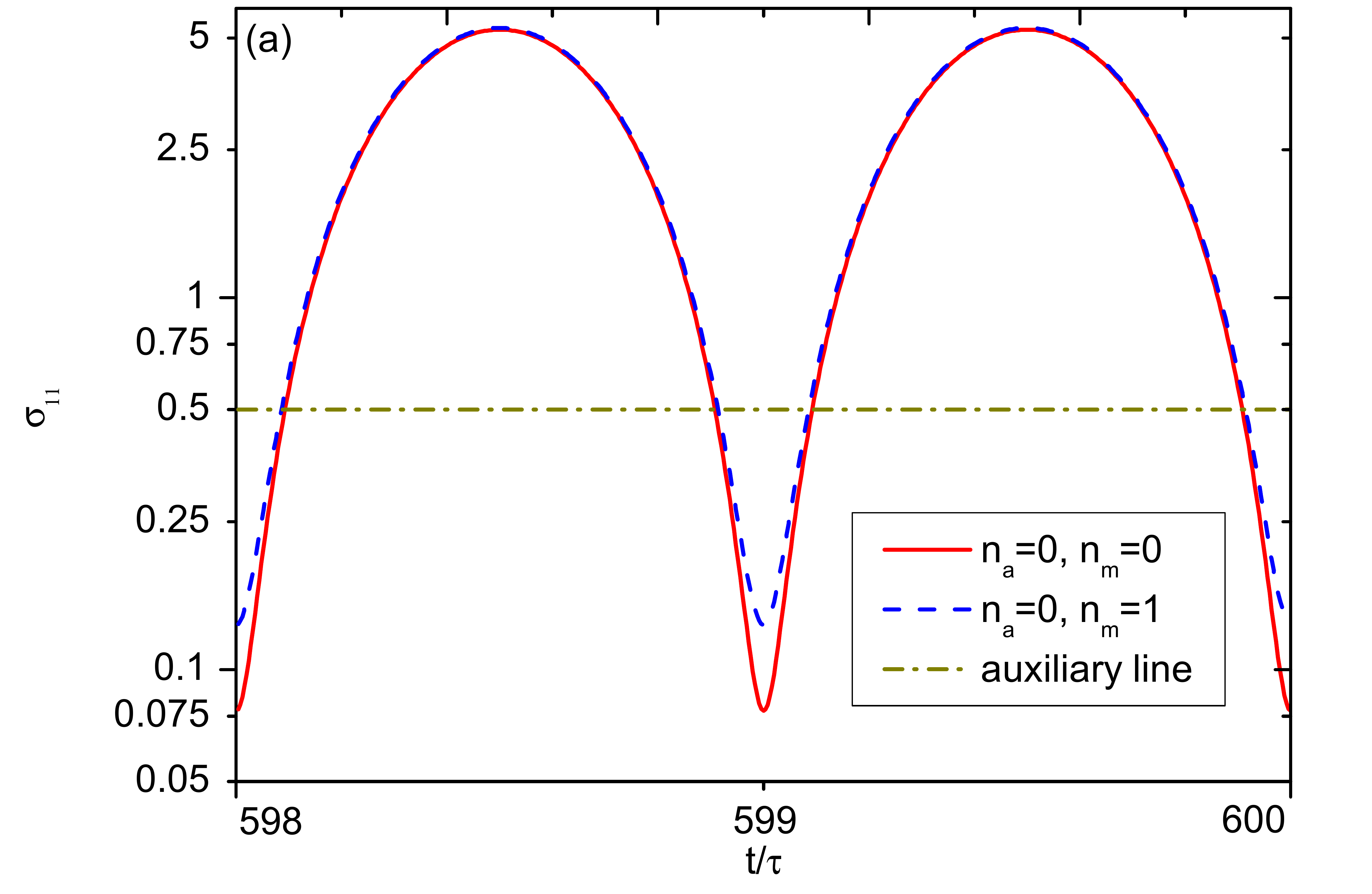}\label{fig:squeezea}}
   \subfigure{\includegraphics[width=0.4\textwidth]{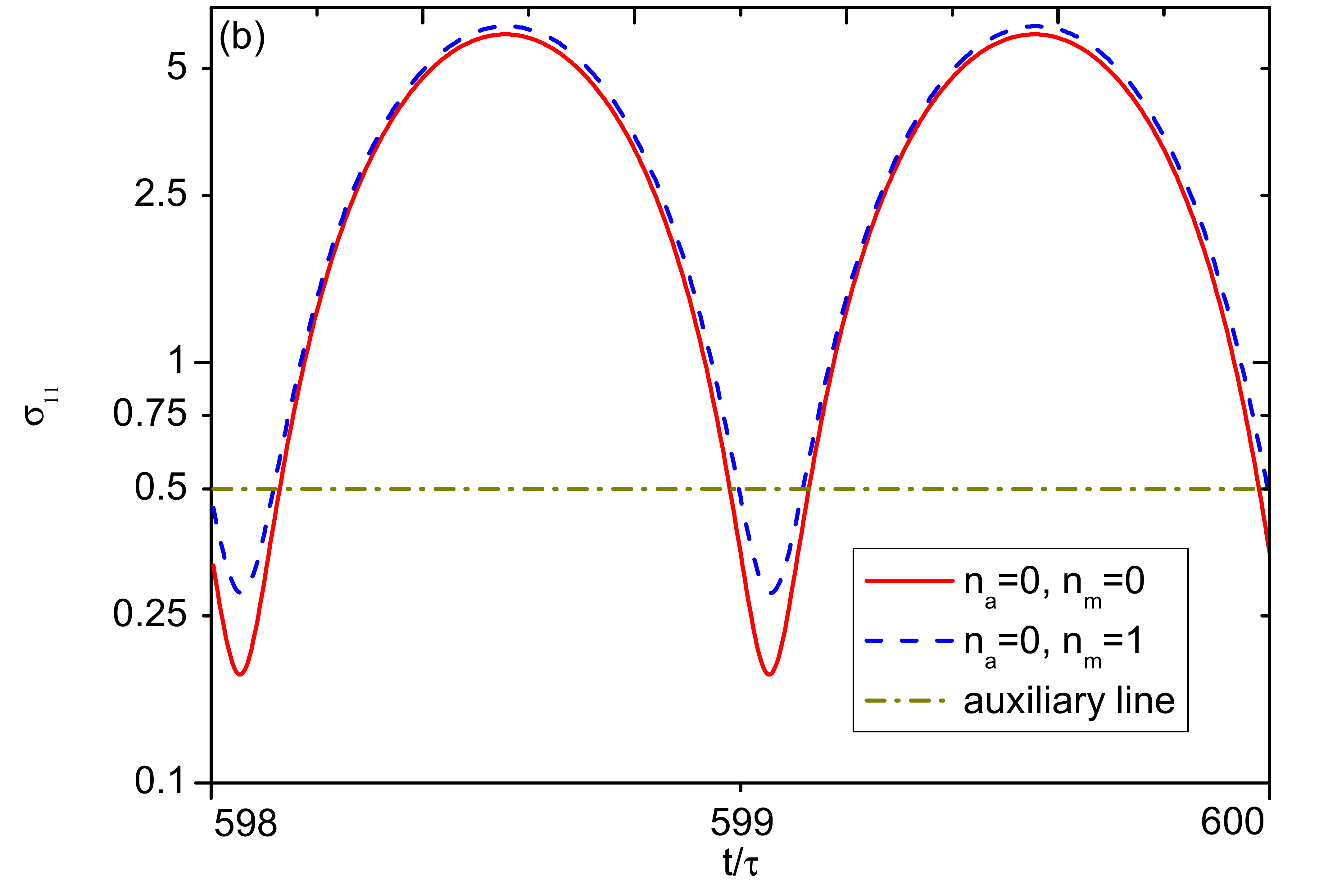}\label{fig:squeezeb}}
   \subfigure{\includegraphics[width=0.4\textwidth]{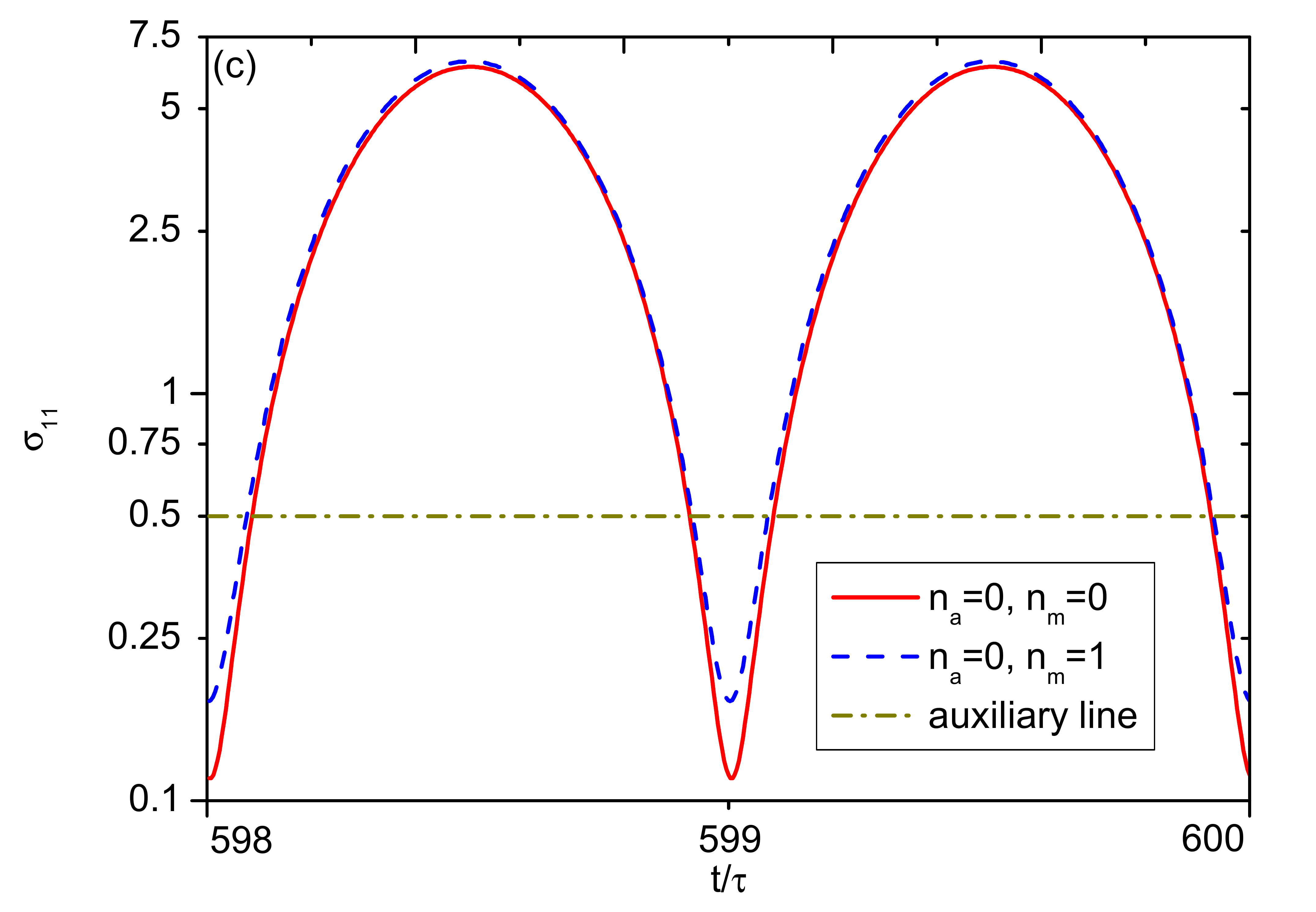}\label{fig:squeezec}}
   \caption{\label{fig:squeeze}
   Variance of the mechanical oscillator position operator ${\sigma _{11}}(t)$ as a function of time in the long time limit from $t = 598\tau $ to $t = 600\tau $. (a)$\Omega  = 2$ for symmetric modulation; (b)$\Omega  = 1.97$ for single cavity driving; (c)$\Omega  = 1.97$ for single cavity modulation. In all figures, the solid (red) and dashed (blue) lines correspond to the cases of ${\overline n _{\rm{a}}} = 0,{\overline n _{\rm{m}}} = 0$ and ${\overline n _{\rm{a}}} = 0,{\overline n _{\rm{m}}} = 1$  respectively and are plotted with logarithmic coordinates. The other parameters are the same as those in Fig.~\ref{fig:mean}.}
\end{figure}

\begin{figure}[ht!]
   \centering
   \subfigure{\includegraphics[width=0.4\textwidth]{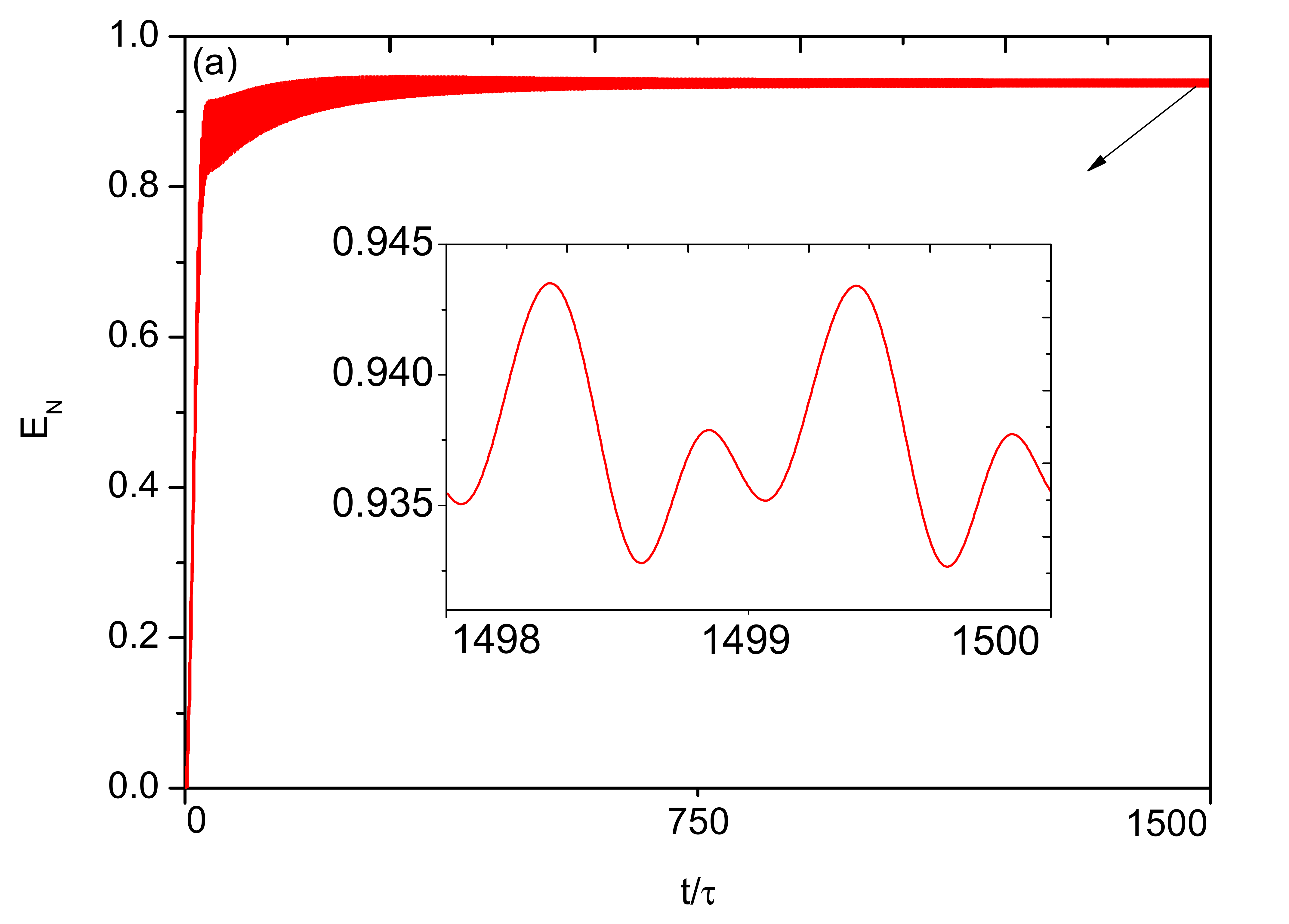}\label{fig:entanglementa}}
   \subfigure{\includegraphics[width=0.4\textwidth]{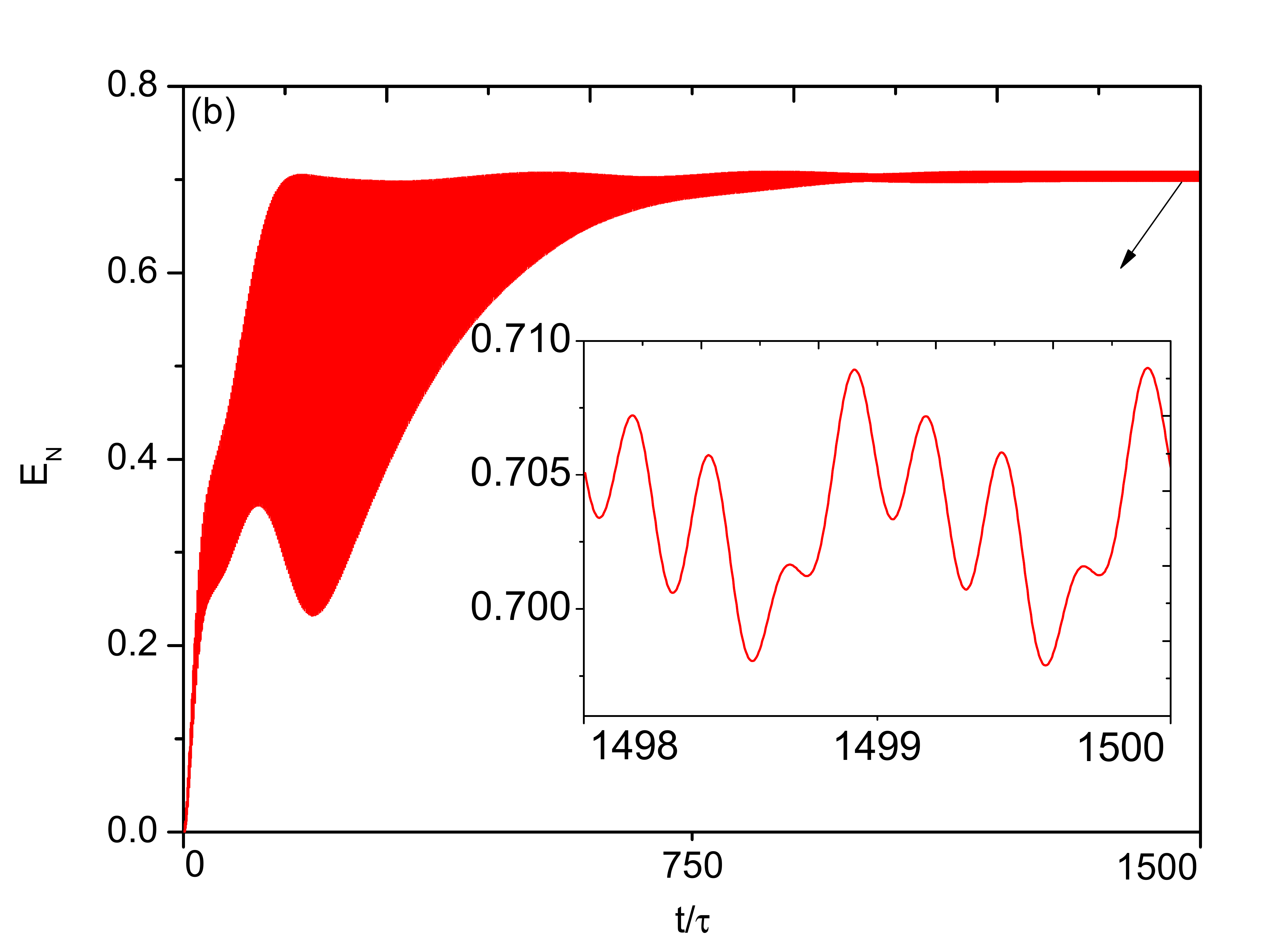}\label{fig:entanglementb}}
   \subfigure{\includegraphics[width=0.4\textwidth]{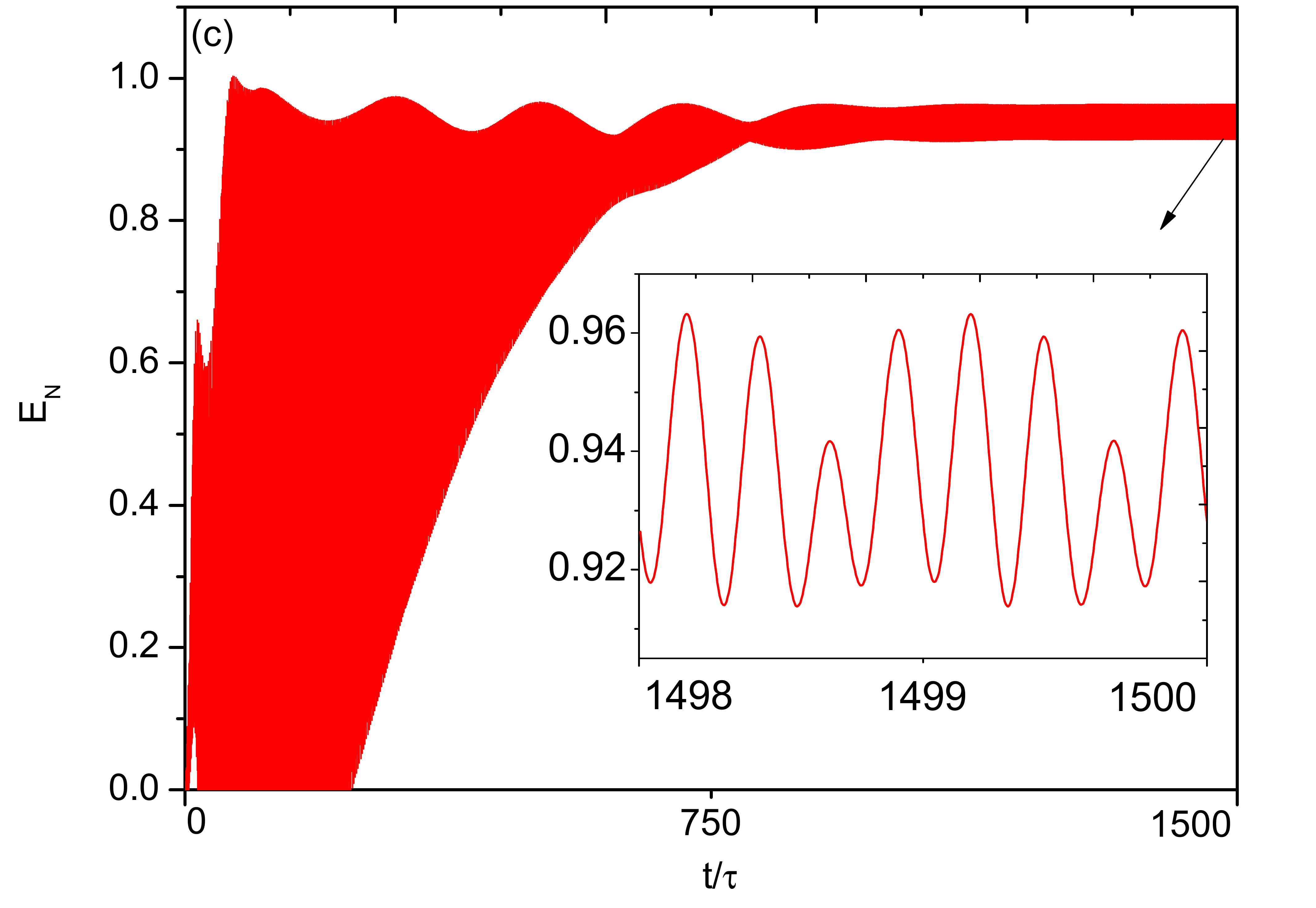}\label{fig:entanglementc}}
   \caption{\label{fig:entanglement}
   Asymptotic evolution of cavity-cavity entanglement ${E_{\rm{N}}}$ as a function of time in the long time limit from $t = 0 $ to $t = 1500\tau $. We take ${\overline n _{\rm{a}}} = 0,{\overline n _{\rm{m}}} = 0$, and $\Omega  = 2$. (a) symmetric modulation; (b) single cavity driving; (c) single cavity modulation. The other parameters are the same as those in Fig.~\ref{fig:mean} except $\kappa  = 0.001$ and ${\gamma _{\rm{m}}} = 0.1$.}
\end{figure}

Figure~\ref{fig:squeeze} displays the asymptotic evolution of the first row and the first column element ${\sigma _{11}}(t)$ of CM, namely variance of the mechanical oscillator position operator, while Fig.~\ref{fig:entanglement} shows the asymptotic evolution of cavity-cavity entanglement ${E_{\rm{N}}}$ for symmetrical and asymmetrical modulations, where all results are only numerically calculated since the numerical results of mean value agree well with the analytical results after hundreds of modulation periods. Here and the following, the stability of the system can be guaranteed by all eigenvalues of the matrix $R(t)$ having a negative real part for all time, which is justified based on the Routh-Hurwitz criterion \cite{dejesus1987routh}. Obviously, the squeezing of the mechanical mode and the cavity-cavity entanglement are indeed $\tau$ period when the system finally tends to be stable in the long time limit. Noticeably, the squeezing and the entanglement can be significantly enhanced compared with the parametric interaction, which are limited by a factor of $1/2$ below the zero-point level, i.e., $0.25$ (the so-called $3 dB$ limit)\cite{Mari2009Gently,woolley2008nanomechanical,nunnenkamp2010cooling,liao2011parametric,schmidt2012optomechanical}, and $0.69$ \cite{paternostro2007creating,Vitali2007Optomechanical,Mari2009Gently,schmidt2012optomechanical}, respectively.

In order to better understanding the physical reality, we introduce the creation and annihilation operators of the mechanical fluctuations

\begin{equation}\label{bn}
b = (q + ip) /\sqrt{2}, b^{\dagger} = (q - ip) /\sqrt{2}
\end{equation}
and the nonlocal bosonic modes
\begin{equation}\label{an}
c_1 = (a_L + a_R) /\sqrt{2}, c_2 = (a_L - a_R) /\sqrt{2}.
\end{equation}
Thus, the linearized system Hamiltonian in Eq.~(\ref{lH}) can be rewritten as
\begin{eqnarray}\label{lH1}
{H^{{\rm{lin}}}}
=& {\Delta _3}c_1^ {\dag} {c_1} + {\Delta _4}c_2^ {\dag} {c_2} + {\omega _{\rm{m}}}{b^ {\dag} }b + \frac{1}{{2\sqrt 2 }}\{ [G_{\rm{L}}^ * (t)- G_{\rm{R}}^ * (t)]{c_1}\\\nonumber
&+ [G_{\rm{L}}^ * (t) + G_{\rm{R}}^ * (t)]{c_2}+ {\rm{h}}{\rm{.c}}{\rm{.}}\} (b + {b^{\dag} }),
\end{eqnarray}
where ${\Delta _3} = {\Delta _1}(t) + J$, ${\Delta _4} = {\Delta _2}(t) - J$.
In the interaction picture with respect to the free part ${\Delta _3}c_1^ {\dag} {c_1} + {\Delta _4}c_2^ {\dag} {c_2} + {\omega _m}{b^ {\dag} }b$, if the relationship between the effective coupling ${G_j}(t)$ and effective mean value of the cavity modes $\left\langle {{A_j}(t)} \right\rangle $ is taken as Eq.~(\ref{effG}), Eq.~(\ref{lH1}) is transformed to
\begin{eqnarray}\label{lH2}
\mathop H\limits^ \sim   &= &\frac{g}{2}\{ [({A_{{\rm{L0}}}} - {A_{{\rm{R0}}}}){e^{ - i({\Delta _3} + {\omega _{\rm{m}}})t}} + ({A_{{\rm{L1}}}} - {A_{{\rm{R1}}}}){e^{ - i({\Delta _3} + {\omega _{\rm{m}}} - \Omega )t}}]{c_1}b \nonumber\\
&&+ [({A_{{\rm{L0}}}} + {A_{{\rm{R0}}}}){e^{ - i({\Delta _4} + {\omega _{\rm{m}}})t}} + ({A_{{\rm{L1}}}} + {A_{{\rm{R1}}}}){e^{ - i({\Delta _4} + {\omega _{\rm{m}}} - \Omega )t}}]{c_2}b\nonumber\\
&&{\rm{      }} + [({A_{{\rm{L0}}}} - {A_{{\rm{R0}}}}){e^{ - i({\Delta _3} - {\omega _{\rm{m}}})t}} + ({A_{{\rm{L1}}}} - {A_{{\rm{R1}}}}){e^{ - i({\Delta _3} - {\omega _{\rm{m}}} - \Omega )t}}]{c_1}{b^ {\dag} }\nonumber\\
&&{\rm{      }} + [({A_{{\rm{L0}}}} + {A_{{\rm{R0}}}}){e^{ - i({\Delta _4} - {\omega _{\rm{m}}})t}} + ({A_{{\rm{L1}}}} + {A_{{\rm{R1}}}}){e^{ - i({\Delta _4} - {\omega _{\rm{m}}} - \Omega )t}}]{c_2}{b^ {\dag} } + {\rm{h}}{\rm{.c}}{\rm{.}}\}.
\end{eqnarray}
Here we focus on the range $g{A_{j0}},g{A_{j1}} \ll {\omega _{\rm{m}}},\Omega $, and set $J = 2{\omega _{\rm{m}}}$, ${\Delta _{\rm{L}}} = {\Delta _{\rm{R}}} = 3{\omega _{\rm{m}}}$ without loss of generality. Based on Eqs.~(\ref{lap}) and~(\ref{effdelta}), we have
\begin{subequations}
\begin{align}
&{\Delta _3} = {\Delta _1}(t) + J \simeq 5{\omega _{\rm{m}}} + \frac{{{g^2}(A_{{\rm{R0}}}^{\rm{2}} + A_{{\rm{R1}}}^{\rm{2}} - A_{{\rm{L0}}}^{\rm{2}} - A_{{\rm{L1}}}^{\rm{2}})}}{{{\omega _{\rm{m}}}}},\\
&{\Delta _4} = {\Delta _2}(t) - J \simeq {\omega _{\rm{m}}} - \frac{{{g^2}(A_{{\rm{R0}}}^2 + A_{{\rm{R}}1}^2 - A_{{\rm{L0}}}^{\rm{2}} - A_{{\rm{L1}}}^{\rm{2}})}}{{{\omega _{\rm{m}}}}},
\end{align}
\end{subequations}
where the fast oscillating terms ${e^{ \pm i\Omega t}}$ have been neglected.
When the modulation frequency is chosen to match with the resonance frequency of the nonlocal cavity and mechanical modes, i.e.,
\begin{equation}\label{omega}
\Omega  = 2{\omega _{\rm{m}}} - \frac{{{g^2}(A_{{\rm{R0}}}^2 + A_{{\rm{R1}}}^2 - A_{{\rm{L0}}}^{\rm{2}} - A_{{\rm{L1}}}^{\rm{2}})}}{{{\omega _{\rm{m}}}}},
\end{equation}
all rapid oscillating terms in Eq.~(\ref{lH2}) can be neglected and the Hamiltonian can be rewritten as
\begin{eqnarray}\label{lH3}
\mathop H\limits^ \sim
\simeq \frac{g}{2}[({A_{{\rm{L1}}}} + {A_{{\rm{R1}}}}){c_2}b + ({A_{{\rm{L0}}}} + {A_{{\rm{R0}}}})\times{e^{\frac{{i{g^2}(A_{{\rm{R}}0}^2 + A_{{\rm{R}}1}^2 - A_{{\rm{L}}0}^2 - A_{{\rm{L}}1}^2)t}}{{{\omega _{\rm{m}}}}}}}{c_2}{b^ {\dag} } + {\rm{h}}{\rm{.c}}{\rm{.}}].
\end{eqnarray}
Due to the fact that $ig^2(A_{\rm{R0}}^2+ A_{\rm{R1}}^2 - A_{\rm{L0}}^2 - A_{\rm{L1}}^2)/\omega_{\rm{m}}\ll A_{\rm{L0}}+A_{\rm{R0}}$, the slow varying term $({A_{{\rm{L0}}}} + {A_{{\rm{R0}}}}){e^{{{i{g^2}(A_{{\rm{R}}0}^2 + A_{{\rm{R}}1}^2 - A_{{\rm{L}}0}^2 - A_{{\rm{L}}1}^2)t} \mathord{\left/
 {\vphantom {{i{g^2}(A_{{\rm{R}}0}^2 + A_{{\rm{R}}1}^2 - A_{{\rm{L}}0}^2 - A_{{\rm{L}}1}^2)t} {{\omega _{\rm{m}}}}}} \right.
 \kern-\nulldelimiterspace} {{\omega _{\rm{m}}}}}}}{c_2}{b^ {\dag} }$ is roughly treated as a costant $({A_{{\rm{L0}}}} + {A_{{\rm{R0}}}}){c_2}{b^ {\dag} }$ for the simplicity in the following analyses.

 Introducing two Bogoliubov-mode annihilation operators
\begin{eqnarray}
&&{\beta _1} = b\cosh r + {b^ {\dag} }\sinh r,\\
&&{\beta _2} = {c_2}\cosh r + c_2^{\dag} \sinh r,
\end{eqnarray}
where the squeezing parameter $r$ is defined as $\tanh r = {{({A_{{\rm{L1}}}} + {A_{{\rm{R1}}}})} \mathord{\left/
 {\vphantom {{({A_{{\rm{L1}}}} + {A_{{\rm{R1}}}})} {({A_{{\rm{L0}}}} + {A_{{\rm{R0}}}})}}} \right.
 \kern-\nulldelimiterspace} {({A_{{\rm{L0}}}} + {A_{{\rm{R0}}}})}}$. Assuming ${A_{{\rm{L1}}}} + {A_{{\rm{R1}}}} < {A_{{\rm{L0}}}} + {A_{{\rm{R0}}}}$, which ensures stability of the system, the Hamiltonian of Eq.~(\ref{lH3}) becomes

\begin{equation}\label{Heffect}
\mathop H\limits^ \sim   \simeq \chi {c_2}\beta _1^ {\dag}  + {\rm{h}}{\rm{.c}}.
\end{equation}
or
\begin{equation}\label{Heffect1}
\mathop H\limits^ \sim   \simeq \chi b\beta _2^ {\dag}  + {\rm{h}}{\rm{.c}}.
\end{equation}
with the coupling
\begin{equation}
\chi  = {{g\sqrt {{{({A_{{\rm{L0}}}} + {A_{{\rm{R0}}}})}^2} - {{({A_{{\rm{L1}}}} + {A_{{\rm{R1}}}})}^2}} } \mathord{\left/
 {\vphantom {{g\sqrt {{{({A_{{\rm{L0}}}} + {A_{{\rm{R0}}}})}^2} - {{({A_{{\rm{L1}}}} + {A_{{\rm{R1}}}})}^2}} } 2}} \right.
 \kern-\nulldelimiterspace} 2}.
\end{equation}
This is a beam-splitter-like Hamiltonian, which is well known from optomechanical sideband cooling \cite{marquardt2007quantum,wilson2007theory}.
Obviously, the ground state of ${\beta _1}$ or ${\beta _2}$ is the single-mode squeezed state of the mechanical mode $b$ or two-mode squeezed state of the cavity modes ${a_{\rm{L}}}$ and ${a_{\rm{R}}}$, respectively.
When the mechanical decay rate $\gamma _{\rm{m}}$ is small, which ensures that the mechanical mode $b$ only weakly couples to the mechanical thermal baths with relatively large mean thermal occupancies, the dynamics of mechanical mode $b$, i.e., the Bogoliubov mode ${\beta _1}$, is dominated by the interaction with the nonlocal bosonic modes ${c_2}$, namely, the cavity modes ${a_{\rm{L}}}$ and ${a_{\rm{R}}}$. Therefore, the Bogoliubov mode ${\beta _1}$ can be cooled to near ground state via the beam-splitter-like interaction [Eq.~(\ref{Heffect})] with the nonlocal bosonic modes ${c_2}$, which strongly interacts with optical thermal baths with neglectable small mean thermal occupancies. In other words, the dissipative dynamics of the cavity modes can be used to cool the Bogoliubov mode ${\beta _1}$, generating single-mode squeezing of the mechanical mode. In contrary, if the cavity decay rate $\kappa$ is smaller compared with the mechanical decay rate $\gamma _{\rm{m}}$ and the mechanical mode $b$ has been precooled by a cold reservoir, as discussed in~\cite{wang2013reservoir}, the beam-splitter-like interaction between the mechanical mode $b$ and the Bogoliubov mode ${\beta _2}$ [Eq.~(\ref{Heffect1})] can be exploited to cool the cavities, obtaining the stationary two-mode squeezing state of two cavities.
The system dynamics behaviors numerically shown in Figs.~\ref{fig:squeeze} and~\ref{fig:entanglement} can be explained very well by the above analyses. Notably, all of the above analyses are based on the assumption that the system is stable and does not enter the chaotic regime \cite{monifi2016optomechanically,lu2015p,Bakemeier2014Route}. Under the circumstance, the amount of stationary squeezing or entanglement is a nonmonotonic function of the ratio of the effective mean value ${{({A_{{\rm{L1}}}} + {A_{{\rm{R1}}}})} \mathord{\left/ {\vphantom {{({A_{{\rm{L1}}}} + {A_{{\rm{R1}}}})} {({A_{{\rm{L0}}}} + {A_{{\rm{R0}}}})}}} \right.
 \kern-\nulldelimiterspace} {({A_{{\rm{L0}}}} + {A_{{\rm{R0}}}})}}$ or the ratio of the effective coupling ${{({G_{{\rm{L1}}}} + {G_{{\rm{R1}}}})} \mathord{\left/ {\vphantom {{({G_{{\rm{L1}}}} + {G_{{\rm{R1}}}})} {({G_{{\rm{L0}}}} + {G_{{\rm{R0}}}})}}} \right.
 \kern-\nulldelimiterspace} {({G_{{\rm{L0}}}} + {G_{{\rm{R0}}}})}}$. According to the Hamiltonian in Eqs.~(\ref{Heffect}) or (\ref{Heffect1}), the increase of the ratio has two competing effects. On the one hand, it can increase the squeezing parameter $r = {{{{\tanh }^{ - 1}}({A_{{\rm{L1}}}} + {A_{{\rm{R1}}}})} \mathord{\left/
 {\vphantom {{{{\tanh }^{ - 1}}({A_{{\rm{L1}}}} + {A_{{\rm{R1}}}})} {({A_{{\rm{L0}}}} + {A_{{\rm{R0}}}})}}} \right.
 \kern-\nulldelimiterspace} {({A_{{\rm{L0}}}} + {A_{{\rm{R0}}}})}}$ and enhance the stationary squeezing and entanglement. On the other hand, it can weaken the cooling effects by declining the coupling strength of the beam-splitter-like interaction. Thus the optimum parameters are a tradeoff between these two competing effects. Accordingly, for a group of specifically optimum parameter values of ${A_{{\rm{L1}}}}$,${A_{{\rm{R1}}}}$, ${A_{{\rm{L0}}}}$ and ${A_{{\rm{R0}}}}$ as Eq.~(\ref{Aas}), the optimum modulations of driving lasers are completely determined by Eqs.~(\ref{lap}) and~(\ref{lap1}), which depend on the parameters $\kappa $, ${\gamma _{\rm{m}}}$, $\Delta_{\rm{L}}$, $\Delta_{\rm{R}}$, $\Omega$, $g$, and $\omega_{\rm{m}}$. Noteworthily, the choice of parameters $\kappa $ and ${\gamma _{\rm{m}}}$ varies with different purpose, resulting in the optimum modulation of driving lasers being also different. The modulation of driving lasers adopted in Figs.~\ref{fig:squeeze} and~\ref{fig:entanglement} may be not the optimal, which is not our focus of concern. Here, we only verify the enhancement of the squeezing and entanglement via symmetrically and asymmetrically periodically modulated lasers. As shown in Fig.~\ref{fig:mean}, when the system is stable, the evolutions of mean value of two cavity modes are fully synchronized in the case of symmetric modulation, leading to the parameters ${A_{{\rm{L1}}}}={A_{{\rm{R1}}}}, {A_{{\rm{L0}}}}={A_{{\rm{R0}}}}$. Under the circumstance of asymmetric modulation, the amplitudes of mean value of two cavity modes are no longer equal, i.e., ${A_{{\rm{L1}}}}\neq{A_{{\rm{R1}}}}, {A_{{\rm{L0}}}}\neq{A_{{\rm{R0}}}}$. However, both symmetric and asymmetric modulations can achieve the same period $\tau$ of the system steady state. The difference is that the needed number of modulation period to achieve stable result varies with modulation mechanisms. Since the amount of stationary squeezing or entanglement depends on the ratio of the effective mean value ${{({A_{{\rm{L1}}}} + {A_{{\rm{R1}}}})} \mathord{\left/ {\vphantom {{({A_{{\rm{L1}}}} + {A_{{\rm{R1}}}})} {({A_{{\rm{L0}}}} + {A_{{\rm{R0}}}})}}} \right.
 \kern-\nulldelimiterspace} {({A_{{\rm{L0}}}} + {A_{{\rm{R0}}}})}}$ rather than the specific value of each parameter, both symmetric and asymmetric modulations of the external driving laser are effective, provided that
the effective mean value ${{({A_{{\rm{L1}}}} + {A_{{\rm{R1}}}})} \mathord{\left/ {\vphantom {{({A_{{\rm{L1}}}} + {A_{{\rm{R1}}}})} {({A_{{\rm{L0}}}} + {A_{{\rm{R0}}}})}}} \right.
 \kern-\nulldelimiterspace} {({A_{{\rm{L0}}}} + {A_{{\rm{R0}}}})}}$ is optimized.

\begin{figure}[t]
\centering
  {\includegraphics[width=0.4\textwidth]{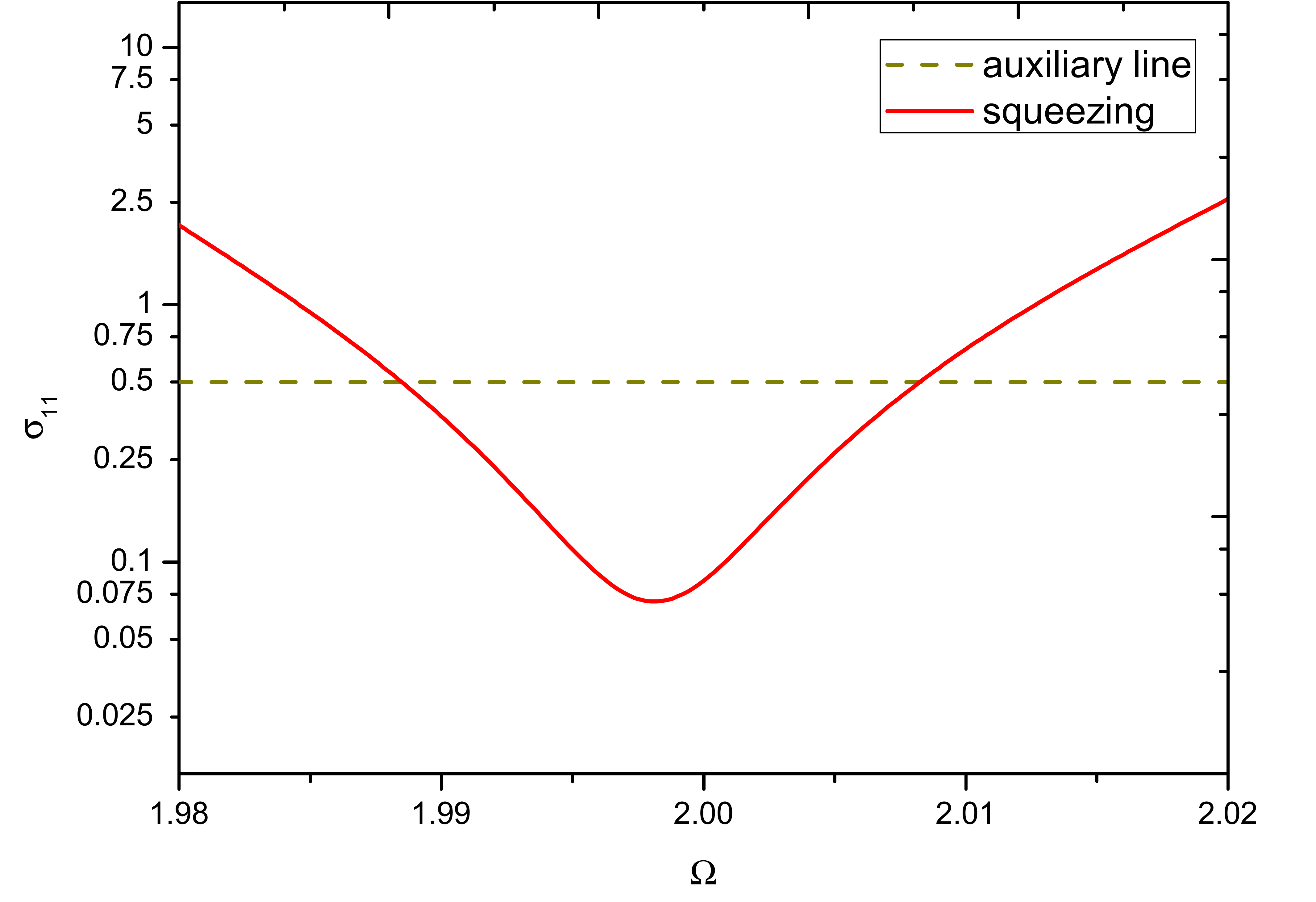}}
  \caption{\label{fig:squeezevia}
  Mimimum variance ${\sigma _{11,\min }}$ of the mechanical oscillator position operator versus the modulation frequency $\Omega $. The chosen parameters in units of ${\omega _{\rm{m}}}$ are $\kappa  = 0.1$, ${\gamma _{\rm{m}}} = 0.001$,$J = 2$, $\Delta  = 3$, $g = 4 \times {10^{ - 6}}$, ${G_{{\rm{L}}0}} = 0.13$, ${G_{{\rm{L}}1}} = 0.12$, ${G_{{\rm{R}}0}} = 0.07$, and ${G_{{\rm{R}}1}} = 0.06$.}
\end{figure}

\begin{figure}[t]
\centering
  {\includegraphics[width=0.4\textwidth]{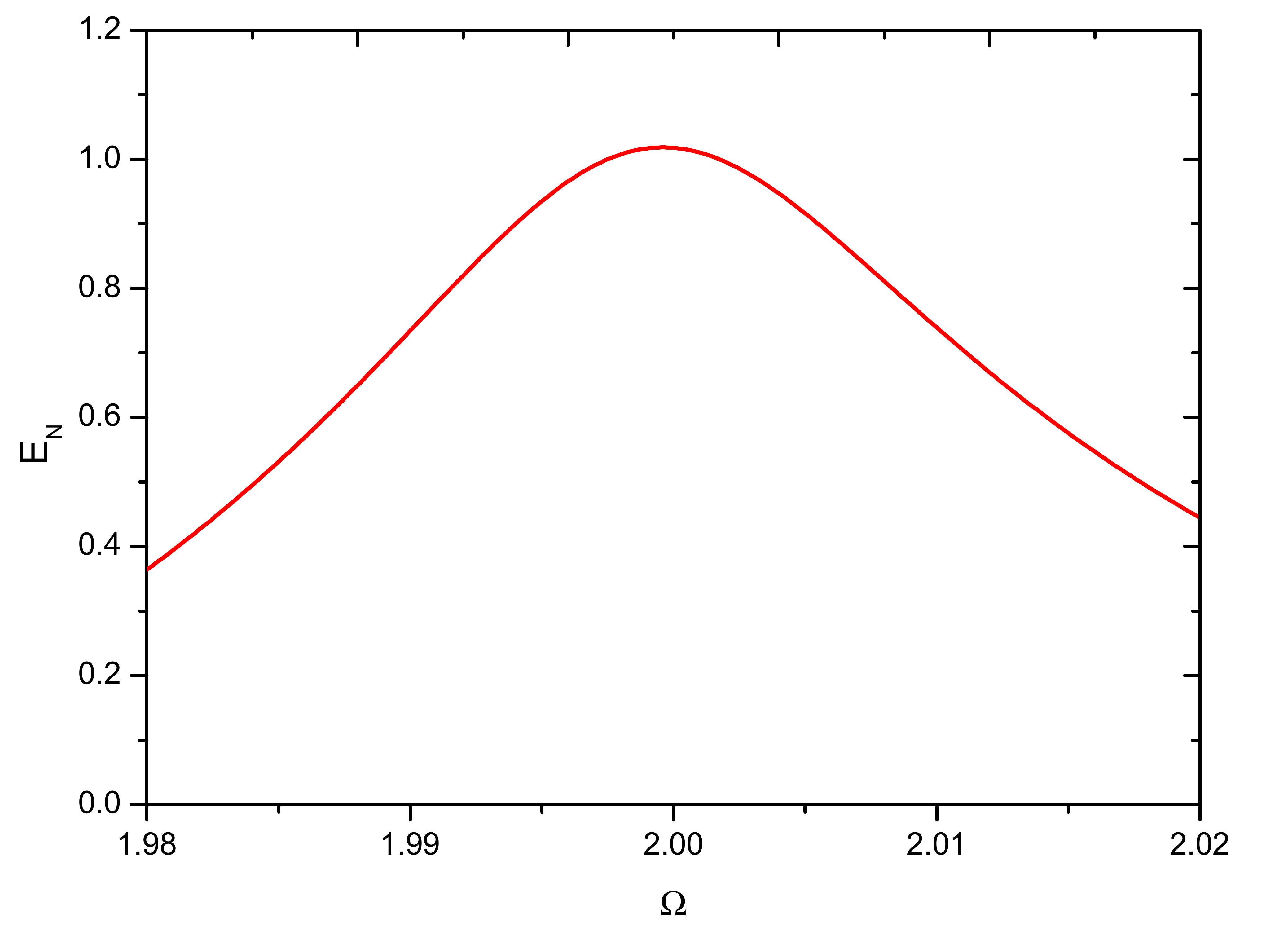}}
  \caption{\label{fig:entanglementvia}
 Maximum cavity-cavity entanglement ${E_{{\rm{N,max}}}}$ versus the modulation frequency $\Omega $. All the other parameters are the same as those in Fig.~\ref{fig:squeezevia} except $\kappa  = 0.001$, ${\gamma _{\rm{m}}} = 0.1$.}
\end{figure}

In order to explore the effect of the optimal modulation frequency on single-mode squeezing and two-mode squeezing, the mimimum variance ${\sigma _{11,\min }}$ of the mechanical oscillator position operator and the maximum cavity-cavity entanglement ${E_{{\rm{N,max}}}}$ as a function of the modulation frequency $\Omega $ are plotted in Figs.~\ref{fig:squeezevia} and \ref{fig:entanglementvia}, respectively. The results are numerically evaluated by applying the corresponding exact mean values in Eq.~(\ref{lap}) with ${\overline n _{\rm{a}}} = 0,{\overline n _{\rm{m}}} = 0$ and the parameters ${G_{{\rm{L}}0}}=0.13$, ${G_{{\rm{L}}1}}=0.12$, ${G_{{\rm{R}}0}}=0.07$, and ${G_{{\rm{R}}1}}=0.06$. Though the group parameters may not be the optimal values, the results of Figs.~\ref{fig:squeezevia} and \ref{fig:entanglementvia} show that the optimal modulation frequency is close to $2\omega_{\rm{m}}$, which is indeed the same as the result predicted by Eq.~(\ref{omega}). Besides, compared to the squeezing of the mechanical oscillator position operator, the cavity-cavity entanglement has a larger scale of modulation frequency $\Omega$, which implies that the squeezing is more sensitive to the variation of the modulation frequency.

\section{Conclusions}
In summary, we have explored the mechanism of periodic driving laser modulation in a dissipative three-mode optomechanical system. Our studies show that combinations of the modulation and the dissipation can significantly enhance the mechanical squeezing and cavity-cavity entanglement. What is more, both symmetric and asymmetric modulations of the external driving laser are effective when we carefully balance the two opposing effects by varying the ratio of the effective mean values of cavity modes or effective coupling. The numerical simulation results signify that it is sufficient to enhance the squeezing and entanglement effects as long as one periodically modulated laser is applied to either end of the cavities, which is convenient for actual experiment. However, the cost is more modulation periods required for achieving system stability. In order to achieve large squeezing and entanglement, apart from selecting appropriate ratio of the effective mean values of cavity modes or effective coupling, the modulation frequency should also be chosen carefully.

\section*{Funding}
National Natural Science Foundation of China (NSFC) (61275215, 11674059); Natural Science Foundation of Fujian Province of China (NSFFPC) (2016J01009, 2013J01008);
Fujian Provincial College Funds for Young and Middle-aged Teacher (FPCFYMT) (JAT160687, JA14397); Fujian Provincial College Funds for Distinguished Young Scientists (FPCFDYS) (JA16); Research Projects of Fujian Polytechnic of Information Technology (RPFPIT) (Y17104).

\section*{Acknowledgments}
We thank Rong-Xin Chen for fruitful discussions.

\end{document}